\global\def\draftcontrol{0}

   \def\versionno{Annulons in Witten's YM}  
  
\catcode`\@=11  
    
\expandafter\ifx\csname draftcontrol\endcsname\relax\global\def\draftcontrol{0}  
\fi  
  
{\count255=\time\divide\count255 by 60  
\xdef\hourmin{\number\count255}  
\multiply\count255 by-60\advance\count255 by\time  
\xdef\hourmin{\hourmin:\ifnum\count255<10 0\fi\the\count255}}  
\def\draftdate{\number\month/\number\day/\number\year\ \ \ \hourmin }

\newcommand\makepapertitle{\par  
  \begingroup \renewcommand\thefootnote{\@fnsymbol\c@footnote}%
    \def\@makefnmark{\rlap{\@textsuperscript{\normalfont\@thefnmark}}}%
    \long\def\@makefntext##1{\parindent 1em\noindent \hb@xt@1.8em{%
    \hss\@textsuperscript{\normalfont\@thefnmark}}##1}
    \global\@topnum\z@   
    \@makepapertitle \thispagestyle{empty}\@thanks \endgroup  
    \setcounter{footnote}{0}
    \global\let\makepapertitle\relax \global\let\@makepapertitle\relax  
    \global\let\@thanks\@empty \global\let\@author\@empty  
    \global\let\@date\@empty \global\let\@title\@empty  
    \global\let\title\relax \global\let\author\relax  
    \global\let\date\relax \global\let\and\relax  
    \def\version{\let\version\@version\@gobble} }  
\def\@makepapertitle{%
  \newpage \ifnum\draftcontrol=1 {} \version\versionno \vskip 3em%
   \else \hfill\hbox to 3cm {\parbox{4cm}{\@pubnum}\hss}
   \fi \begin{center}
   \vskip 1.5em
   \begin{tabular}[t]{c}
   {\@bstract}
}  
  
\gdef\@pubnum{}  
\def\pubnum#1{%
  \gdef\@pubnum{#1}}  
  
\gdef\@bstract{}  
\def\Abstract#1{%
  \gdef\@bstract{%
   \parbox{\textwidth-0pc}{%
   \centerline{\bf Abstract}\penalty1000%
\noindent
\renewcommand\baselinestretch{1.0}%
{#1}}}}

\def\ps@paper{\let\@mkboth\@gobbletwo%
     \ifnum\draftcontrol=1  
        \def\@oddfoot{\hbox to \textwidth{\tiny \versionno \hfil\tiny\draftdate}%
        \hskip -\textwidth \hbox to \textwidth{\hfil\rm\thepage\hfil}}%
     \else\def\@oddfoot{\hbox to \textwidth{\hfil\rm\thepage\hfil}}  
     \fi  
     \let\@evenfoot\@oddfoot  
}

\def\@version#1{\ifnum\draftcontrol=1  
\typeout{}\typeout{#1}\typeout{}  
\vskip3mm\centerline{\hbox{\fbox{\normalsize{\tt DRAFT -- #1 -- }  
                   {\draftdate}}}}\vskip3mm  
\fi}  
\let\version\@version  
\long\def\eqlabel#1{\ifnum\draftcontrol=1  
                    \tag@false  
                    \tag*{(\theequation) \hbox to -0.2cm{\hspace{0cm}\small{#1}\hss}}  
                    \refstepcounter{equation}  
                    \edef\@currentlabel{\theequation}  
                    \ltx@label{#1}          
                    \else  
                    \label{#1}  
                    \fi  
                    }  
\let\st@bibitem\@bibitem  
\let\st@lbibitem\@lbibitem  
\ifnum\draftcontrol=1  
  \def\@bibitem#1{%
    \st@bibitem{#1}\a@@label{#1}\ignorespaces}  
  \def\@lbibitem[#1]#2{%
    \st@lbibitem[#1]{#2}\a@@label{#2}\ignorespaces}  
  \def\a@@label#1{%
    \gdef\a@lab{\smash{\normalfont\small#1}}  
    \ifvmode  
      \if@inlabel  
        \global\setbox\@labels\hbox{%
          \llap{\a@lab\let\a@lab\relax  
                \kern\@totalleftmargin\kern\marginparsep}%
          \box\@labels}%
      \fi  
    \fi}  
\fi  

\documentclass[12pt,letterpaper]{article}  

\usepackage{amsmath,amssymb,array,calc,epsfig,latexsym}  
\ifnum\draftcontrol=1  
\fi  
  
\renewcommand\baselinestretch{1.25}  
\renewcommand\section{\@startsection {section}{1}{\z@}%
                                   {-3.5ex \@plus -1ex \@minus -.2ex}%
                                   {2.3ex \@plus.2ex}%
                                   {\normalfont\large\bfseries}}  
\renewcommand\subsection{\@startsection{subsection}{2}{\z@}%
                                   {-3.25ex\@plus -1ex \@minus -.2ex}%
                                   {1.5ex \@plus .2ex}%
                                   {\normalfont\normalsize\bfseries}}  
\renewcommand\subsubsection{\@startsection{subsubsection}{3}{\z@}%
                                   {-3.25ex\@plus -1ex \@minus -.2ex}%
                                   {1.5ex \@plus .2ex}%
                                   {\normalfont\normalsize\it}}  
\renewcommand\paragraph{\@startsection{paragraph}{4}{\z@}%
                                   {-3.25ex\@plus -1ex \@minus -.2ex}%
                                   {1.5ex \@plus .2ex}%
                                   {\normalfont\normalsize\bf}}  
  
\def\revise#1       {\raisebox{-0em}{\rule{3pt}{1em}}%
                     \marginpar{\raisebox{.5em}{\vrule width3pt\  
                     \vrule width0pt height 0pt depth0.5em \hbox to  
                     0cm{\hspace{0cm}{%
                     \parbox[t]{4em}{\raggedright\footnotesize{#1}}}\hss}}}}  
  
\def\ul{\underline}  

\def\caln         {{\cal N}}

\def\del          {\partial}

\def\tr           {\mathop{\rm Tr}}

\def\half{{\frac12}}

\def\sqr#1#2{{\vcenter{\vbox{\hrule height.#2pt  
 \hbox{\vrule width.#2pt height#1pt \kern#1pt \vrule width.#2pt}\hrule  
 height.#2pt}}}}

\def\a{\alpha}  
\def\b{\beta}  
\def\r{\rho}

\def\g{\gamma}

\catcode`\@=12  
  
\begin{document}  

  

\newcommand{\be}{\begin{equation}}  
\newcommand{\ee}{\end{equation}}  
\newcommand{\beq}{\begin{equation}}  
\newcommand{\eeq}{\end{equation}}  
\newcommand{\ba}{\begin{eqnarray}}  
\newcommand{\ea}{\end{eqnarray}}  
\newcommand{\nn}{\nonumber}  

\def\vol{\bf vol}  
\def\Vol{\bf Vol}  
\def\del{{\partial}}  
\def\vev#1{\left\langle #1 \right\rangle}  
\def\cn{{\cal N}}  
\def\co{{\cal O}}  
\def\IC{{\mathbb C}}  
\def\IR{{\mathbb R}}  
\def\IZ{{\mathbb Z}}  
\def\RP{{\bf RP}}  
\def\CP{{\bf CP}}  
\def\Poincare{{Poincar\'e }}  
\def\tr{{\rm tr}}  
\def\tp{{\tilde \Phi}}  
\def\Y{{\bf Y}}  
\def\te{\theta}  
\def\bX{\bf{X}}  
  
\def\TL{\hfil$\displaystyle{##}$}  
\def\TR{$\displaystyle{{}##}$\hfil}  
\def\TC{\hfil$\displaystyle{##}$\hfil}  
\def\TT{\hbox{##}}  
\def\HLINE{\noalign{\vskip1\jot}\hline\noalign{\vskip1\jot}} 
\def\seqalign#1#2{\vcenter{\openup1\jot  
  \halign{\strut #1\cr #2 \cr}}}  
\def\lbldef#1#2{\expandafter\gdef\csname #1\endcsname {#2}}  
\def\eqn#1#2{\lbldef{#1}{(\ref{#1})}%
\begin{equation} #2 \label{#1} \end{equation}}  
\def\eqalign#1{\vcenter{\openup1\jot  
    \halign{\strut\span\TL & \span\TR\cr #1 \cr }}}  
\def\eno#1{(\ref{#1})}  
\def\href#1#2{#2}  
\def\half{{1 \over 2}}

\def\ads{{\it AdS}}  
\def\adsp{{\it AdS}$_{p+2}$}  
\def\cft{{\it CFT}}  
  
\newcommand{\ber}{\begin{eqnarray}}  
\newcommand{\eer}{\end{eqnarray}}  
  
\newcommand{\bea}{\begin{eqnarray}}  
\newcommand{\eea}{\end{eqnarray}}

\newcommand{\beqar}{\begin{eqnarray}}  
\newcommand{\cN}{{\cal N}}  
\newcommand{\cO}{{\cal O}}  
\newcommand{\cA}{{\cal A}}  
\newcommand{\cT}{{\cal T}}  
\newcommand{\cF}{{\cal F}}  
\newcommand{\cC}{{\cal C}}  
\newcommand{\cR}{{\cal R}}  
\newcommand{\cW}{{\cal W}}  
\newcommand{\eeqar}{\end{eqnarray}}  
\newcommand{\lm}{\lambda}\newcommand{\Lm}{\Lambda}  
\newcommand{\eps}{\epsilon}  
  

\newcommand{\nonu}{\nonumber}  
\newcommand{\oh}{\displaystyle{\frac{1}{2}}}  
\newcommand{\dsl}  
  {\kern.06em\hbox{\raise.15ex\hbox{$/$}\kern-.56em\hbox{$\partial$}}}  
\newcommand{\as}{\not\!\! A}  
\newcommand{\ps}{\not\! p}  
\newcommand{\ks}{\not\! k}  
\newcommand{\D}{{\cal{D}}}  
\newcommand{\dv}{d^2x}  
\newcommand{\Z}{{\cal Z}}  
\newcommand{\N}{{\cal N}}  
\newcommand{\Dsl}{\not\!\! D}  
\newcommand{\Bsl}{\not\!\! B}  
\newcommand{\Psl}{\not\!\! P}  
\newcommand{\eeqarr}{\end{eqnarray}}  
\newcommand{\ZZ}{{\rm \kern 0.275em Z \kern -0.92em Z}\;}

\def\s{\sigma}  
\def\a{\alpha}  
\def\b{\beta}  
\def\r{\rho}  
\def\d{\delta}  
\def\g{\gamma}  
\def\G{\Gamma}  
\def\ep{\epsilon}  
\makeatletter \@addtoreset{equation}{section} \makeatother  
\renewcommand{\theequation}{\thesection.\arabic{equation}}  

\begin{titlepage}  
  
\version\versionno  
  
\leftline{\tt hep-th/0409205}  
  
\vskip -0.5cm

\rightline{\small{\tt MCTP-04-40}} 
\vskip -0.2cm  
\rightline{\small{\tt CPHT-RR-041-0704}}  
\vskip -0.2cm
\rightline{\small{\tt IC/2004/53}} 
\vskip -0.2cm
\rightline{\small{\tt IFUM-806-FT}}  

\vskip .7 cm  
  
 \centerline{\bf \Large Wilson Loop, Regge Trajectory and Hadron Masses}
\vskip .4cm

\centerline{\bf \Large in a Yang-Mills Theory from Semiclassical Strings}
\vskip .8cm

 \centerline{\large  F. Bigazzi${}^1$, A. L.  Cotrone${}^{2}$, L. Martucci$^{3}$ and  L. A. Pando Zayas${}^{4}$}


\vskip .8cm \centerline{\it ${}^1$ The Abdus Salam ICTP,
Strada Costiera, 11; I-34014 Trieste, Italy.}

\vskip .2cm \centerline{\it ${}^2$ CPHT,  \'Ecole Polytechnique;
F-91128 Palaiseau Cedex, France.}

\vskip .2cm \centerline{\it ${}^2$ INFN,  Piazza dei Caprettari, 70;
I-00186  Roma, Italy.}

\vskip .2cm \centerline{\it ${}^3$ Dipartimento di Fisica,
Universit\'a di Milano, Via Celoria, 16; I-20133 Milano, Italy.}

\vskip .2cm \centerline{\it ${}^4$ Michigan Center for Theoretical
Physics, Randall Laboratory of Physics,} \centerline{\it The
University of Michigan, Ann Arbor, MI 48109-1120.  USA}


\begin{abstract}  
We compute the one-loop string corrections to the  Wilson loop, glueball Regge
trajectory and stringy hadron masses in the Witten model
of non supersymmetric, large-$N$ Yang-Mills theory.  
The classical string configurations corresponding to the above
field theory objects are respectively: open straight strings, folded closed spinning
strings, 
and  strings orbiting in the internal part of the supergravity background. 
For the rectangular Wilson loop we show that besides the standard
L\"uscher term, string corrections provide a rescaling of the field theory
string tension. The
one-loop corrections to the linear glueball Regge trajectories render
them nonlinear with 
a positive
intercept, as in the experimental soft
Pomeron trajectory.
Strings orbiting in the internal
space predict a spectrum of
hadronic-like states charged under global flavor symmetries which
falls in the same universality class of other confining models.
\end{abstract}  
  
\end{titlepage}

\tableofcontents   

\section{Introduction}
One of the most remarkable developments in the holographic approach to
solving gauge theories has been the realization of certain scenarios
where the correspondence can be taken beyond the supergravity
approximation \cite{bmn,gkp}. The main idea behind this development
asserts that some classical configurations of the string sigma model
in a given supergravity background are dual, in the holographic sense,
to states of the gauge theory. In particular,  the conserved
quantities of the classical configurations are to be identified with
the quantum numbers describing gauge theory states.

Being ultimately interested in realistic gauge theories, a natural
question in this approach pertains to the properties of states of
confining gauge theories.  Unlike the case of the standard duality
between strings in $AdS_5\times S^5$ and $\caln=4$ SYM, for the case of
confining strings perturbative field theory calculations are extremely
difficult.  The only data available to test the predictions of the
holographic approach come from lattice simulations of $SU(N_c)$ gauge
theories and some low energy phenomenology of QCD. Evidently, 
holographic calculations are  valid for large $N_c$ and a direct
comparison with QCD is still beyond our current reach. Nevertheless,
it is appropriate to compare the general behavior on both sides. For
example, a recent holographic analysis of corrections to the glueball
Regge trajectories seems to be compatible with experimental data
describing the soft Pomeron trajectory \cite{regge}. Other studies
point to a possible emergence of the chiral perturbation Lagrangian
from a holographic point of view \cite{cpth}.

In the search for a holographic dual to  nonsupersymmetric Yang-Mills
one is faced with the hunting task of producing a holographic
background which ideally retains some of the control provided by
supersymmetric theories. Several examples of soft breaking of
$\caln=1$ backgrounds have been provided \cite{softbreaking,son},
however, in this process one usually looses analytic control over the
supergravity background  if the scale of the breaking is high enough
for the study of YM theory. A different route to pure YM was proposed
by Witten in \cite{wqcd} and consists of a stack of D4-branes wrapped
on a thermal circle. This implies that on the gauge theory side we
deal with a 5-d gauge theory compactified on a circle along which the
fermions have antiperiodic boundary conditions. For the appropriate
energy scale this theory is essentially nonsupersymmetric YM in 4-d.

In this paper we present a detailed study of the one-loop sigma-model
corrections to various classical string solutions, in the asymptotic
region of the dual supergravity background  conjectured to be relevant
for the description of the IR regime of the field theory.

The paper is organized as follows.    In section
\ref{cla} we present the background and the string theory action needed for the
computation of the one-loop corrections. 

For supergravity backgrounds dual to confining gauge theories the
classical string describing a rectangular Wilson loop with  
area-law behavior goes into the bulk with a bathtub-shaped
configuration. 
In section \ref{wil} we consider the  open string configuration that
approximates this Wilson loop sufficiently well: an open
straight string of length $L$ lying at the minimal transverse radius
of the dual background. 
We then calculate the leading sigma-model correction to its energy.
There are other evaluations in the Witten model of the one-loop
quantum correction to this classical configuration
\cite{rey,olesen,sonne}. A distinctive aspect of our calculation is the
systematic treatment of the fermionic sector and that we provide a
consistency check via cancellation of anomalies. 
Our calculation  provides  a
stringy prediction for the leading quantum correction to the linear
quark anti-quark potential in a theory ``close'' to pure, large $N$,
YM. 
We find that the dominant term in this correction is not of L\"uscher
($1/L$) type, but amounts to a kind of ``renormalization'' of the effective string tension\footnote{We call it a ``renormalization'' throughout the paper for brevity. It is really a correction to the dependence of the tension on the UV coupling $\lambda$, rather than a dependence on the running coupling.}. 
We find a L\"uscher-like term, with negative sign, as a subleading correction. 

In section \ref{reg} we consider a closed string at the minimal radius
spinning in the (warped) flat 4-d directions of the background. In the
dual gauge theory this configuration describes glueball Regge
trajectories. 
Even if we are not able to fully solve the string spectral problem for this configuration, 
we show that in some interesting regimes
the one-loop corrected glueball Regge
trajectory becomes non-linear with a positive intercept. 
These
features are qualitatively shared (through in a different regime) by the best fit to the experimental
soft Pomeron trajectory\footnote{For an interesting recent study of the glueball Regge trajectories on the lattice and 
in a ``old'' string model of QCD, see \cite{meyer}.} given by the UA8 collaboration \cite{ua8}. 
We also find, again, a ``renormalization'' of the string tension.

In section \ref{int} we consider closed multi-spinning string states
rotating on the internal $S^4$ of the IR region of the supergravity 
background. These are argued to correspond to multi-charged hadronic
states in the dual field theory, whose constituents are massive
adjoint fields\footnote{For a study of baryonic spectra from strings in $AdS$ space see \cite{brod}.}. We consider the classical energy in the large spin
limit and discuss  the quantum corrections for the simplest
circular solutions. Their behavior is expected to be qualitatively
equivalent to the one of the  multi-spinning string solutions examined
in the literature both in the confining and in the non confining cases. In
particular, in the semiclassical limit of large total spin $J$, the
leading quantum correction to the energy reduces to the one
corresponding to collapsed point-like strings with spin $J$ orbiting
along the equator of $S^4$. The one-loop corrections in this case can be simply
deduced by quantizing closed strings on the plane wave background obtained
from the Penrose limit of the Witten model. The corresponding string
spectrum is studied in section \ref{ppw}. It shares the universal
properties of all the other confining examples examined in literature
\cite{gpss,alfra,son,gfacl}. The string results give thus the expected
universal predictions on the energy spectrum of multi-charged hadronic
states, called Annulons, in gauge theory.

We conclude in section \ref{com} with comments on the stringy
predictions for the dual field theory and the nature of their
universality. 

We include also several appendices with some technical details.
In appendix \ref{revw} we
briefly  review the supergravity background paying special attention
to the regime of validity of the supergravity description and to aspects
of the holographic relations with the dual gauge theory.
In appendix \ref{appsol} we discuss the general form of the classical
solutions at constant radius.
Appendix \ref{appappr} shows how the straight open string approximates well the
bathtub-shaped configuration describing the actual Wilson loop, in the
large $L$ limit, both from the classical and semi-classical point of view. 
We discuss the cancellation of the Weyl anomaly for the Regge trajectory configuration
in appendix \ref{appweyl}.
Finally, appendix \ref{apppenr} provides details on the Penrose limit of the Witten background.

The supergravity background we consider in this paper has recently
received renewed attention, many of the classical configurations we
analyze are also discussed in \cite{petkou,ali,pons}. Our main
contribution is a systematic description  of the one-loop corrections
to such classical configurations.

\section{General setup} \label{cla} 

In this section we introduce the Witten background, we rewrite the metric in the relevant IR regime (formulae (\ref{IRmetric}) and (\ref{rmetric})) and present
the bosonic and fermionic string actions we will use in the following in (\ref{bosac}) and (\ref{feracflat}).
More details of the supergravity background, with some notes on the holographic relations with the field theory, are collected
in appendix \ref{revw}.

The ten-dimensional string frame metric and
dilaton of the Witten model are given by 
\bea  
\label{defns} 
ds^2&=&({u\over
R})^{3/2} (\eta_{\mu\nu}dx^\mu dx^\nu + {4R^3\over
9u_0}f(u)d\theta^2)+ ({R\over u})^{3/2}{du^2\over f(u)}
+R^{3/2}u^{1/2}d\Omega_4^2\ ,\nonumber \\    
f(u)&=&1-{u_0^3\over
u^3}\ , \qquad \qquad \qquad R=(\pi
Ng_s)^{1\over3}{\alpha'}^{1\over2}\ , \nonumber \\    
e^\Phi&=&g_s{u^{3/4}\over R^{3/4}}\ . 
\eea
The geometry consists of a warped, flat 4-d part, a radial direction $u$, a circle parameterized by $\theta$ with radius vanishing at the horizon $u=u_0$, and a four-sphere whose volume is instead everywhere non-zero.
It is non-singular at $u=u_0$.
Notice that in the $u\to\infty$
limit the dilaton diverges: this implies  that in this limit the
completion of the present IIA model has to be found in M-theory.
The background is completed by a constant four-form field strength  
\be
F_4=3R^3\omega_4\ ,  
\ee    
where $\omega_4$ is the volume form of the
transverse $S^4$.  

The main gauge theory parameter we will use in the following is the KK mass scale
$1/R_{\theta}$, which is given by  
\bea \label{emmezero}
\frac{1}{R_\theta}=\frac32m_0\ ,\qquad {\rm where}\qquad
m_0^2=\frac{u_0}{R^3}\ .    
\eea    
As can be read from the metric,
$m_0$ is also the typical glueball mass scale and, as we will show in
the following (see formula (\ref{areal})), its square is proportional to the ratio between the
confining string tension $T_{QCD}$ and the UV 't Hooft coupling
$\lambda$.  
As usual, the supergravity approximation is reliable  in the regime
opposite  to that in which the KK degrees of freedom can decouple from
the low energy dynamics.
The condition $T_{QCD}\ll m_0^2$ implies in fact $\lambda\ll 1$, which
is beyond the supergravity regime of validity.

We will be mainly interested in classical string configurations
localized at the horizon $u=u_0$, since this region is dual to the IR regime of the dual field theory.
In this case the
coordinate $u$ is not suitable because the metric written in
this coordinate looks  singular at $u=u_0$. Then, as a first step, let
us  introduce the radial coordinate 
\bea 
r^2=\frac{u-u_0}{u_0}\ , 
\eea
so that the metric expanded to quadratic order around $r=0$ becomes
\be
\label{IRmetric}   
ds^2\approx ({u_0\over R})^{3/2}[1+{3r^2\over
2}](\eta_{\mu\nu}dx^\mu dx^\nu) + {4\over3}R^{3/2}\sqrt{u_0}(dr^2+ r^2
d\theta^2) + R^{3/2}u_0^{1/2}[1+ {r^2\over 2}]d\Omega_4^2\ .  
\ee 

In order to
simplify the study of the classical string configurations and their
semiclassical quantization, it is useful to rescale the dimensional
coordinates with the KK mass parameter which represents the reference
scale of our theory. We do it by defining the dimensionless
coordinates $X^\mu=m_0 x^\mu$. Introducing the parameter
$\xi=\lambda/3$, the metric in the IR   
takes the form \bea\label{rmetric}   ds^2=l_s^2\xi \Big\{
\big[1+\frac32(y_1^2+y_2^2)\big] dX^\mu
dX_\mu+\frac43(dy_1^2+dy_2^2)+[1+\frac12(y_1^2+y_2^2)]d\Omega_4\Big\}\
, \eea   where we have used Cartesian  coordinates $y_a,\ a=1,2$, such
that $r^2=y_1^2+y_2^2$, and the metric is expanded up to the second
order in $y^2_a$.  From (\ref{rmetric}) one can easily see that the
string action contains only $\xi$ as external parameter.

\subsection{Action}\label{genactions}

As a first step we are going to consider classical solutions sitting at the
origin $y^a=0$ of the $y^a$'s plane
for which we can use the following effective Polyakov action in the
conformal gauge \bea   S=-\frac{\xi}{4\pi}\int d\tau d\sigma
\big[\partial_\alpha X^\mu\partial^\alpha X_\mu +\partial_\alpha
\zeta^A\partial^\alpha \zeta^A -\Lambda (\zeta^A\zeta^A-1)\big]\ ,
\eea   where the Lagrange multiplier $\Lambda$ constraints $\zeta^A,\
A=1,\ldots,5,$  to define the $S^4$, i.e. $\zeta^A\zeta^A=1$.
  
We want to classify the solutions by choosing $J^{12}$ and $J^{34}$,
i.e. the generators of the rotations in the 1-2 and 3-4 planes of the
coordinates $\zeta^A$, as a basis of the Cartan sub-algebra of the
isometry group $SO(5)$.  A useful parameterization of $S^4$ is the
following: \bea   \zeta^5=\sin\psi\ ,\qquad \zeta^M=\cos\psi Z^M\
,\quad M=1,\ldots,4\ ,\quad {\rm with}\quad Z^MZ^M=1\ .     \eea Then,
the $S^4$ metric takes the form \bea d\Omega_4^2=d\psi^2+\cos^2\psi
d\Omega_3^2\ .  \eea   If we consider solutions located at the point
$\zeta^5=0\Rightarrow \psi=0$\footnote{In this way, we are looking for
classical configurations corresponding to highest weight states of
$so(5)$ with respect to the Cartan sub-algebra defined by $J^{12}$ and
$J^{34}$.}, we can take as effective action \bea\label{bosac}
S=-\frac{\xi}{4\pi}\int d\tau d\sigma \big[\partial_\alpha
X^\mu\partial^\alpha X_\mu  +\partial_\alpha Z^M\partial^\alpha Z^M
-\Lambda (Z^MZ^M-1)\big]\ .  \eea

The other ingredient we need is the type IIA GS action expanded up to
the second order in the fermions. In his  Polyakov form, the quadratic
fermionic action is \cite{ms} 
\bea
\label{faction}
S_F&=&\frac{i}{4\pi\alpha^\prime}\int d\tau d\sigma
\sqrt{-h}\bar\theta(1+\Gamma_{F1})\Gamma^\alpha D_\alpha\theta=\cr
&=&\frac{i}{4\pi\alpha^\prime}\int d\tau d\sigma
\bar\theta[(\sqrt{-h}h^{\alpha\beta}+\epsilon^{\alpha\beta}\Gamma^{\ul{11}})\Gamma_\alpha
D_\beta]\theta  \ , 
\eea   where $\theta(\tau,\sigma)$ is a ten dimensional Majorana spinor, $\Gamma_\alpha$ is the pull-back on
the string world-sheet of $\Gamma_m=e^{\ul{a}}_m\Gamma_{\ul a}$ ($m\ ({\ul{a}})$ are the generic
curved (flat) ten dimensional indexes\footnote{In order to avoid ambiguities, we underline
flat indexes.}), $D_m$ is the usual generalized
covariant derivative entering the supersymmetry transformations of the
gravitino in type IIA supergravity, and \bea
\Gamma_{F1}=\frac{\epsilon^{\alpha\beta}}{2\sqrt{-h}}\Gamma_{\alpha\beta}\Gamma^{\ul{11}}
\eea   is the natural chiral world-sheet operator. The kappa-symmetry
acts on the $\theta$ in the following way \bea
\delta_{\kappa}\theta=(1-\Gamma_{F1})\kappa\ ,   \eea    where
$\kappa$ is an arbitrary Majorana spinor. In our case where only the
RR $F_{(4)}$ is turned on, the generalized covariant derivative
reduces to 
\bea   
D_m=\partial_m+\frac14
\omega_{m{\ul{ab}}}\Gamma^{\ul{ab}}-\frac{1}{8\cdot 4!\,g_s}e^\Phi
F^{(4)}_{\ul{abcd}}\Gamma^{\ul{abcd}}\Gamma_m\ .  
\eea
The configurations we will consider can be grouped into two kinds of
solutions both satisfying the static gauge condition $X^0\equiv t
=k\tau$ and either moving in the $\mathbb{R}^{1,3}$ parameterized by
the $X^\mu$'s  or rotating on the $S^4$.
Let us then introduce the following  natural  vielbein for the IR region of
our background (\ref{rmetric}) 
\bea   &&
e^{\ul{\mu}}=\sqrt{\xi\alpha^\prime}dX^\mu\ ,\quad \mu=0,\ldots,3\ ,
\\  && e^{\ul{4}}=\sqrt{\frac43\xi\alpha^\prime}dy^1\ ,\qquad \ 
e^{\ul{5}}=\sqrt{\frac43\xi\alpha^\prime}dy^2\ ,\qquad \qquad \quad
e^{\ul{6}}=\sqrt{\xi\alpha^\prime}d\psi\ ,\cr &&
e^{\ul{7}}=\sqrt{\xi\alpha^\prime}\cos\psi d\chi\ ,\quad
e^{\ul{8}}=\sqrt{\xi\alpha^\prime}\cos\psi\sin\chi d\phi_-\ ,\quad
e^{\ul{9}}=\sqrt{\xi\alpha^\prime}\cos\psi\cos\chi d\phi_+\ .\nonumber
\eea   In this basis the RR field-strength is given by 
\bea
F_{\ul{6789}}=\frac{3R^3}{\alpha^{\prime 2}\xi^2}\ .  
\eea   If we
focus on the study of fluctuations around bosonic configurations with
a non-trivial shape $X^\mu(\tau,\sigma)$ only in the flat directions
(as for the Wilson loop and the Regge trajectory in the following),
the pull-back on the world-sheet of the spin-connection vanishes and
the quadratic fermionic action (\ref{faction}) in the conformal gauge,
and after a rescaling\footnote{This is the natural rescaling in order
to see the $\theta$ as the fermions associated to the bosons $X^\mu$
in the canonical way $X^\mu\sim \bar\theta\Gamma^{\ul{\mu}}\theta$.}
$\theta\rightarrow (\alpha^\prime \xi)^{1/4}\theta$,  reduces to \bea
\label{feracflat} S_F&=&\frac{i\xi}{4\pi}\int d\tau d\sigma \Big[
\partial_\alpha
X^\mu\bar\theta(\eta^{\alpha\beta}+\epsilon^{\alpha\beta}\Gamma^{\ul{11}})\Gamma_{\ul{\mu}}\partial_\beta
\theta-\frac38\partial_\alpha X^\mu \partial^\alpha
X_\mu\bar\theta\tilde\Gamma \theta+\cr
&&\quad\quad-\frac38\epsilon^{\alpha\beta}\partial_\alpha X^\mu
\partial_\beta
X^\nu\bar\theta\Gamma^{\ul{11}}\Gamma_{\ul{\mu\nu}}\tilde\Gamma \theta
\Big]\ ,  \eea   where $\tilde\Gamma=\Gamma_{\ul{6789}}$.   In order
to write the kappa-symmetry in the most transparent way, it is useful
to go to a reference frame $\hat e_{\ul a}$ where the first two directions
$\hat e_{\ul{\alpha}}$ are tangent to the world-sheet \bea   \hat
e_{\ul{\alpha}}=\frac{\partial_\alpha X^\mu}{\sqrt{\partial_\sigma
X^\nu\partial_\sigma X_\nu}}e_{\ul{\mu}}\ .   \eea   Now, if we introduce new $\hat\Gamma_{\ul a}$'s 
associated to the vielbein $\hat e^{\ul a}_m$ ($\hat e^{\ul a}_m\hat\Gamma_{\ul a}= e^{\ul a}_m\Gamma_{\ul a}$) and split $\theta=\theta^1+\theta^2$,
where $\theta^I$ ($I=1,2$) have opposite ten dimensional chirality
$\Gamma^{\ul{11}}\theta^I=(-)^{I+1}$, the kappa-symmetry takes the
form \bea   \delta_{\kappa}\theta^1=(1-\hat\Gamma_{\ul{01}})\kappa\
,\qquad\delta_{\kappa}\theta^2=(1+\hat\Gamma_{\ul{01}})\kappa\ .  \eea
Then, in order to fix the kappa-symmetry, it is clear that the most
natural gauge is given by the conditions \bea\label{kfixing} \left\{
\begin{array}{c}  
(1-\hat\Gamma_{\ul{01}})\theta^1=0 \\
(1+\hat\Gamma_{\ul{01}})\theta^2=0
\end{array}\right.  
\quad \Leftrightarrow \quad   \left\{
\begin{array}{c}  
\hat\Gamma^-\theta^1=0 \\   \hat\Gamma^+\theta^2=0
\end{array}\right.\ ,  
\eea   with
$\hat\Gamma^\pm=\hat\Gamma^{\ul{0}}\pm\hat\Gamma^{\ul{1}}$. Even if
transparent from the geometrical point of view, this kappa-fixing
depends in general from $\tau$ and $\sigma$ and requires some attention. An explicit example in which 
we will use a $\sigma$ dependent kappa-fixing  will be considered in section \ref{reg}.

\section{Correction to the rectangular Wilson Loop} \label{wil} 
In this section we consider the quadratic fluctuations around the
classical string configuration corresponding to the gauge theory
rectangular Wilson loop.  The quantization around the true string
solution for the Wilson loop is extremely complicated,  and in the 
literature it is always performed around an approximated
configuration, namely the straight string.  This approximation turns
out to be quite reliable, the corrections being exponentially
suppressed in the regime of interest.
We show in appendix \ref{appappr} how it accurately
approximates the actual solution for the Wilson loop, at both the classical and semi-classical level.  

In the following subsection we
will calculate the bosonic quadratic fluctuations for the straight string sitting at the horizon $u=u_0$ (or
$y^a=0$ in the coordinates used in (\ref{rmetric})).
We then compute the fermionic fluctuations and the quantum correction in
\ref{fqf} and \ref{olc}.
The resulting quantum corrected energy in formula (\ref{eloopfin}) will be discussed in section \ref{willoop}. 

\subsection{Straight string at minimal radius: bosonic fluctuations}\label{wbqf}  

The configuration we are going to study is that of a straight open  
string of length $L$ lying at $y^a=0$ in our
background (\ref{rmetric}) 
\be  
\label{solmi} 
X^0=\tau\ ,\qquad X^1=\sigma\ ,\quad
\sigma\in[-\frac{L}{2},\frac{L}{2}]\ .  
\ee 
The bosonic action
(\ref{bosac}) is very simple in this case and the evaluation of the
energy of the configuration is straightforward.  Its classical value
(per unit time) satisfies the area-law\footnote{Throughout the paper
  we often pass from dimensionless 
quantities to dimensional ones.
The context should help in avoiding confusion.} 
\be 
E = T_{QCD}L\ , \qquad
T_{QCD}={u_0^{3/2}\over 2\pi \alpha'R^{3/2}}=\frac{1}{6\pi}\lambda
m_0^2\ ,
\label{areal}
\ee 
signaling that the dual gauge theory is confining.

The bosonic quadratic fluctuations around this solution are trivial
for the modes on the four-sphere and for the flat 4-d part, as is
transparent from the metric (\ref{rmetric}) and from the fact that the
world-sheet volume factor is equal to one.  The fluctuations of these
modes give just six free massless bosons $\Phi^i,\ i=1,...,6$. 
The configuration has to
satisfy Dirichlet boundary conditions at $\sigma = \pm L/2$, then
\be
\label{solboswilson}   
\Phi^i={\cal N}\sum_n a_n
e^{i\omega_n\tau}\sin{\bigl[\frac{n \pi}{L}\bigl(\sigma+\frac
    L2\bigr)\bigr]}\ , 
\ee   
where ${\cal N}$ is a normalization and $a_n$ are constants.  The associated frequencies are 
\be
\omega_n=\frac{\pi n}{L}\ .   
\ee   
The non-trivial part of the
action is the one for the two $y^{1,2}$ modes.    From (\ref{rmetric})
it follows that they get a mass term from the coupling with the
on-shell $\partial_{\alpha} X^\mu \partial^{\alpha} X_\mu$ part.   The
action for the fluctuation of these modes, expanded in inverse powers
of $\sqrt{\xi}$ around the classical value $y^{1,2}=0$,  reads \be
S=-\frac{1}{4\pi}\int d\tau d\sigma \sum_{a=1,2}\big[\frac43
\partial_\alpha y^a\partial^\alpha y_a + 3y^ay_a\big]\ .  \ee   Upon
canonically normalizing the kinetic term and considering the solutions
for $y^a$ as in (\ref{solboswilson}), the two frequencies read (after
reintroducing the dependence on $R,u_0$) \be
\omega_n=\sqrt{\frac{\pi^2 n^2}{L^2}+\frac{9}{4}m_0^2}\ , \qquad
m_0^2=\frac{u_0}{R^3}\ .
\label{wilbos1}
\ee
In appendix \ref{appappr} it is shown that these frequencies correctly 
approximate the ones for the actual Wilson loop in the large $L$ limit; 
if the string does not sit at the horizon, we only have to replace $u_0$ with $u_m$,
the minimal value of the radius reached by the configuration.

\subsection{Fermionic quadratic fluctuations} \label{fqf} 
In order to study the fermionic contribution to the one-loop
correction to the Wilson loop, it is sufficient to consider the
fermionic action (\ref{feracflat}) expanded around the  minimal radius
configuration (\ref{solmi}).

Following the general setting of section \ref{genactions}, in this case $\hat e^{\ul a}=e^{\ul a}$ and the (kappa-unfixed) fermionic action
takes the form \bea   S_F&=& \frac{i}{2\pi}\int d\tau d\sigma\Big[
\bar\theta^1\Gamma^+\partial_+ \theta^1+\bar\theta^2\Gamma^-\partial_-
\theta^2
-\frac38(\bar\theta^1\tilde\Gamma\theta^2+\bar\theta^2\tilde\Gamma\theta^1)+\cr
&&\quad\quad
+\frac38(\bar\theta^1\Gamma_{\ul{01}}\tilde\Gamma\theta^2-\bar\theta^2\Gamma_{\ul{01}}\tilde\Gamma\theta^1)\Big]\
,  \eea 
where $\partial_\pm =(1/2)(\partial_\tau \pm \partial_\sigma)$.  The kappa-fixing (\ref{kfixing}) can be rendered explicit by
introducing the following set of gamma matrices $\Gamma_{\ul a}$: \bea
\label{gammamat} \Gamma_{\ul{0}}=i\sigma_2\otimes \mathbb{I}\ ,\qquad
\Gamma_{\ul{1}}=\sigma_1\otimes \mathbb{I}\ ,\qquad
\Gamma_{\ul{A}}=\sigma_3\otimes \gamma_{\ul A} \quad(A=2,\dots,9)\ ,   \eea where
$\gamma_{\ul A}$ are Euclidean Dirac matrices in eight dimensions and we can
split the two $\theta^I$ each into two Euclidean 8-d Majorana-Weyl
fermions of opposite chirality. Then, the kappa-fixing conditions
(\ref{kfixing}) becomes $\sigma_3\theta^I=(-)^{I+1}\theta^I$ and we
can write \bea
\label{spinors8d} \theta^1=\frac1{\sqrt2}\left(\begin{array}{c}
\Theta^1 \\ 0\end{array}\right)\ ,\qquad
\theta^2=\frac1{\sqrt2}\left(\begin{array}{c} 0\\ \Theta^2
\end{array}\right)\ ,   \eea     where $\Theta^I$ are two Euclidean
8-d Majorana-Weyl spinor of the same chirality (with respect to the
8-d  chirality operator $\gamma^{\ul{11}}\equiv
\gamma_{\ul 2}\cdots\gamma_{\ul 9}$). Then, the kappa-fixed quadratic fermionic
action becomes \bea   S_F=-\frac{i}{2\pi}\int d\tau d\sigma \Big[
\Theta^1\partial_+\Theta^1 +\Theta^2\partial_-\Theta^2
-\frac38(\Theta^1\tilde\gamma\Theta^2
-\Theta^2\tilde\gamma\Theta^1)\Big]\ , \eea   where
$\tilde\gamma=\gamma_{\ul{6789}}$.      From this action the following
(squared) equation of motion follow for $\Theta^I$ \be
(\partial^2_\tau-\partial^2_\sigma+\frac{9}{16})\Theta^I=0\ .  \ee 
Thus we find eight massive fermionic
modes whose frequency reads (we reinsert here the dependence on $m_0$)
\be \omega_n=\sqrt{\frac{\pi^2n^2}{L^2}+\frac{9}{16}m_0^2}\ .
\label{wilfer1}
\ee  In the non minimal radius case we have simply to substitute
$m_0^2$ with $u_m/R^3$.

\subsection{The one loop correction to the energy}\label{olc}

Let us collect the results found in the previous subsections.
The effective one-loop sigma-model describing the quadratic
fluctuations of the string around the classical solutions is given by
the following collection of free modes: six massless and two massive
bosonic modes, and eight massive fermions. The corresponding non
trivial frequencies are given in (\ref{wilbos1}) and (\ref{wilfer1}).

The one-loop correction to the classical energy $E_c=T_{QCD}L$ thus
reads \be   
E_1=
\frac{\pi}{2L}\sum_{n\geq 1}\left[6\sqrt{n^2}+2\sqrt{n^2+\frac{9m_0^2L^2}{4\pi^2}}-8\sqrt{n^2+\frac{9m_0^2L^2}{16\pi^2}}\right]\
.   
\ee   
This is, as usual \cite{alfra3,alfra2,gfacl}, a negative
function of the effective masses, and hence of $L$.  Let us notice
that the straight string configuration gives a finite UV theory
(without Weyl anomaly \cite{grosstsey})\footnote{The cancellation of
  this divergence provides us with a consistency check for our
  calculation of the frequency of the fluctuations. This check is the
  more crucial due to the various conflicting results appearing in the literature \cite{rey,olesen,sonne}.}
, since 
\be  
\label{weyl}
\sum_{bosons} \omega^2_{bosons} - \sum_{fermions} \omega^2_{fermions}
= 0\ .   \ee   
Thus  $E_1$ is not divergent and in the large $L$
limit it reads 
\be   
E_1\approx -\frac{9m_0^2L}{8\pi}\log{2} -{\pi\over4L} .
\ee   
The leading term in this expression comes from approximating the series above by integrals. The L\"uscher-like term $\pi/4L$ comes from the subleading contribution of the six massless modes. The remaining subleading terms are exponentially suppressed in the large $L$ limit and are related to the massive modes. See \cite{alfra2,gfacl} for details on the evaluation of series like the one above.  

In the large $L$ limit we find that the energy of
the string configuration corresponding to the Wilson loop is given, up
to the one-loop sigma model correction, by 
\be\label{eloopfin}
E\approx E_c + E_1 = {m_0^2\lambda\over6\pi}\Big(1-
{27\over4\lambda}\log2\Big)L -{\pi\over4L}=T_{QCD}\Big(1- {27\over4\lambda}\log2\Big)L-{\pi\over4L}  \ .  \ee
We see that, at the level of approximations we are taking, string
theory gives a prediction on the way the YM string tension
``renormalizes'', $T^{(ren)}_{QCD}(\lambda)=(1- {27\over4\lambda}\log2)T_{QCD}$. 
We will
discuss more extensively these results in section \ref{com}.

\section{Folded closed spinning string: Glueball Regge trajectory} \label{reg}
In this section we consider the closed string configuration corresponding to
the glueball Regge trajectories.  
The relevant closed  folded spinning
string configuration is described, in the coordinates used in
 (\ref{rmetric}), by the following  one parameter family of
solutions \bea\label{reggesolu} X^0=k\tau\ ,\qquad
X^1=k\cos\tau\sin\sigma\ ,\qquad X^2=k\sin\tau\sin\sigma\ , \eea  and
all the other coordinates fixed.

Following standard calculations \cite{regge} and reintroducing the dimensional energy, it is easy to find that
\be
E={g_{00}(u_0)\over \alpha^\prime}k={\lambda\,m_0\over3}k\ , \qquad  J= {g_{00}(u_0)\over2\alpha^\prime}k^2={\lambda\over6}k^2\ .
\label{ejcla}
\ee
Thus the relation between the energy $E$ and the angular momentum $J$ of
the string (in  the $12$ plane) has indeed the expected Regge-like
form \be E^2 = \frac{2}{\alpha'} \left({u_0\over R}\right)^{3/2}J = {{2\over3}\lambda\,m_0^2}J=
4\pi T_{QCD}J\ .  \ee 
Note that the tension of the adjoint string $T_{adj}$ (the one relevant for the Regge trajectories of adjoint particles we are considering) is given by the relation $E^2=J/\alpha'_{adj}=T_{adj}J/2\pi$, i.e. $T_{adj}=2T_{QCD}$.
This is not surprising in field theory since $T_{adj}$ is expected to be related to the string tension between particles in the fundamental, $T_{QCD}$, by the relation
\bea \label{fourfour}
\frac{T_{adj}}{T_{QCD}}=\frac{C_{adj}}{C_{fund}}=\frac{2N^2}{N^2-1}\ ,
\eea
which in the large $N$ limit reduces to the relation above.

In order to obtain the one-loop quantum corrections to these classical
results, we have to study  the quadratic fluctuations around the
classical string configuration (\ref{reggesolu}).  The problem is that
in this case the induced metric is the non-constant  conformally flat
metric  
\bea 
h_{\alpha\beta}=k^2\cos^2\sigma\, \eta_{\alpha\beta}.
\eea 
We will see that as a result the quadratic action has
$\sigma$-dependent external terms coupled to the fluctuating
fields. The one-loop correction  to the classical energy are  more
complicated than in the other cases  considered in this paper.  For
this reason, after computing the bosonic and fermionic quadratic
fluctuations around \ref{rqf}, we calculate the correction to the
Regge trajectory in the limits $k  \ll 1$ in \ref{kmin} and $k  \gg 1$
in \ref{kmag}; the final results in the two cases, presented in 
(\ref{rt}) and (\ref{renregge},\ref{trenreg}) respectively, will be
also discussed in section \ref{regcom}.  We study the finiteness of
the theory in appendix \ref{appweyl}.

\subsection{Quadratic fluctuations}\label{rqf}

By expanding the bosonic world-sheet  fields around the classical
solution (\ref{reggesolu}) one finds that all the fluctuating fields
are free except  two $y_a$'s, which have action  
\bea
\label{actionb}
S=-\frac{1}{4\pi}\int d\tau d\sigma
\sum_{a=1,2}\big[\frac43(\partial_\alpha y_a\partial^\alpha y_a)
+3k^2\cos^2\sigma (y_a)^2\big]\ .  
\eea 
The equations of motion are naturally written in the form
\bea
\big[\partial_\alpha\partial^\alpha -\frac{9k^2}{4}\cos^2\sigma\big]  y_a=0\ .
\eea 
Thus the spectrum of the bosonic fluctuations around the classical
configuration on a flat world-sheet consists of six massless and two massive  modes with an
effective $\sigma$-dependent mass parameter 
\be\label{reggefb} m_B^2 =
\frac{9}{4}k^2\cos^2\sigma \equiv k^2 \, \ell_B^2 \cos^2 \sigma\ .  
\ee
This is qualitatively similar to what happens for the bosonic fluctuations
around the classical string solution corresponding to the Regge trajectory
in other confining backgrounds \cite{regge}. The only difference is in
the number of massive fluctuations which is three for the
Klebanov-Strassler \cite{KS} and Maldacena-N\'u\~nez \cite{mn} backgrounds while in our
case we have only two massive bosonic fluctuations.

Let us now consider the fermionic sector. From the general discussion of
section \ref{genactions} we  know  we can start with  the
action (\ref{feracflat}). As we have explained, the most natural $\kappa$-fixing is given by \bea
(1-\hat\Gamma_{\ul{01}})\theta^1=0\ ,\qquad
(1+\hat\Gamma_{\ul{01}})\theta^2=0\ .  \eea This gauge fixing is
clearly world-sheet coordinate dependent. In order to simplify this
situation, one could proceed in two ways. One could choose a rotating
vielbein (in order to have $\hat \Gamma_{\ul a}$ as constant matrices) or,
more conveniently, one can rotate directly the spinors. In particular,
if the rotation $M(\tau,\sigma)_{\ul a}{}^{\ul b}$ connecting the two vielbeins
(i.e. $\hat e_{\ul a}=M(\tau,\sigma)_{\ul a}{}^{\ul b} e_{\ul b}$) is implemented on the
spinors by $\Lambda(\tau,\sigma)$ (i.e. $\Lambda \Gamma_{\ul a}
\Lambda^{-1}=M_{\ul a}{}^{\ul b}\Gamma_{\ul b}=\hat\Gamma_{\ul a}$), we can introduce  the
rotated spinors $\tilde \theta^I=\Lambda^{-1}\theta^I$ and the
$\kappa$-fixing becomes
\bea
\label{kfixing2} 
(1-\Gamma_{\ul{01}})\tilde\theta^1=0\ , \qquad
(1+\Gamma_{\ul{01}})\tilde\theta^2=0\ ,  \eea   which can be solved
following the Wilson loop example. Since, after an obvious  constant
rescaling of the metric, on the world-sheet the first two one-forms of
the adapted  (co)vielbein are given by  \bea  \hat
e_{\ul\alpha}=\frac{\partial_\alpha X^\mu}{\sqrt{\partial_\sigma
X^\mu\partial_\sigma X_\mu}}\partial_\mu\ , \eea we can choose the
following rotation generator  \bea
\Lambda=e^{\frac\pi4[1-{\rm
sign}(\cos\sigma)]\Gamma_{\ul{12}}}e^{-\frac12 \tau\Gamma_{\ul{12}}} e^{\frac12
\chi(\sigma)\Gamma_{\ul{02}}}\ , \eea where\footnote{Notice that this
transformation is degenerate at the ends of the folded string
$\sigma=\pi/2,3\pi/2$.}
$\chi(\sigma)=\cosh^{-1}\frac{1}{|\cos\sigma|}=\sinh^{-1}\frac{\sin\sigma}{\cos\sigma}$.
The action (\ref{feracflat}) can be written as   \bea
S_F&=&\frac{i}{4\pi}\int d\tau d\sigma \Big[
k|\cos\sigma|\bar\theta(\eta^{\alpha\beta}+\epsilon^{\alpha\beta}\Gamma^{\ul{11}})\hat\Gamma_{\ul{\alpha}}\partial_\beta
\theta-\frac34 k^2\cos^2\sigma\bar\theta\tilde{\Gamma} \theta+\cr
&&\quad\quad-\frac34
k^2\cos^2\sigma\bar\theta\Gamma^{\ul{11}}\hat\Gamma_{\ul{01}}\tilde\Gamma
\theta \Big]\ .  \eea where
$\tilde\Gamma=\hat\Gamma_{\ul{6789}}=\Gamma_{\ul{6789}}$ and
$\theta=\theta^1+\theta^2$. Then, in terms of the rotated spinors it
becomes  \bea  S_F&=&\frac{i}{4\pi}\int d\tau d\sigma \Big[
k|\cos\sigma|\bar{\tilde\theta}(\eta^{\alpha\beta}+\epsilon^{\alpha\beta}\Gamma^{\ul{11}})\Gamma_{\ul{\alpha}}\partial_\beta
\tilde\theta-\frac34 k^2\cos^2\sigma\bar{\tilde\theta}\tilde{\Gamma}
\tilde\theta + \\ &&\quad\quad-\frac34
k^2\cos^2\sigma\bar{\tilde\theta}\Gamma^{\ul{11}}\Gamma_{\ul{01}}\tilde\Gamma
\tilde\theta
+k|\cos\sigma|\bar{\tilde\theta}(\eta^{\alpha\beta}+\epsilon^{\alpha\beta}\Gamma^{\ul{11}})\Gamma_{\ul{\alpha}}(\Lambda^{-1}\partial_\beta\Lambda)
\tilde\theta \Big]\ .\nonumber \eea  Using the fact that the
Majorana-Weyl spinors $\tilde\theta^1$ and $\tilde\theta^2$ have
opposite space-time chirality and satisfy the $\kappa$-fixing
conditions (\ref{kfixing2}), which can be read as a condition of
opposite ``world-sheet chirality'' for the two spinors, the last term
of the action reduces to \bea
&&k|\cos\sigma|\bar{\tilde\theta}(\eta^{\alpha\beta}+\epsilon^{\alpha\beta}\Gamma^{\ul{11}})\Gamma_{\ul{\alpha}}(\Lambda^{-1}\partial_\beta\Lambda)
\tilde\theta \cr 
&& =-\frac k2 {\rm sign}(\cos\sigma) \sin\sigma \bar{\tilde\theta}(\Gamma_{\ul1}+\Gamma^{\ul{11}}\Gamma_{\ul0})\tilde\theta + \cr
&& \quad -\frac {k\pi}{4} \sin\sigma {\rm sign}(\cos\sigma) \partial_\sigma[{\rm sign}(\cos\sigma)]\bar{\tilde\theta}(\Gamma_{\ul0}+\Gamma^{\ul{11}}\Gamma_{\ul1})\tilde\theta \cr
&& =k{\rm sign}(\cos\sigma)\sin\sigma\big[\bar{\tilde\theta}^1\Gamma_{\ul1}\tilde\theta^1+\bar{\tilde\theta}^2\Gamma_{\ul1}\tilde\theta^2
\big]\ .  \eea Indeed, any $\Gamma_{\ul a}$ changes the space-time chirality
while only the two  $\Gamma_{\ul\alpha}$'s ($\alpha=0,1$) change the
world-sheet chirality. Then the above result follows from the fact
that $\Lambda^{-1}\partial_\tau\Lambda= -\frac12(\cosh
\chi\Gamma_{\ul{12}}+\sinh\chi\Gamma_{\ul{01}})$ and
$\Lambda^{-1}\partial_\sigma\Lambda=-\frac {\pi}{4}\partial_\sigma[{\rm sign}(\cos\sigma)] (\cosh
\chi\Gamma_{\ul{12}}+\sinh\chi\Gamma_{\ul{01}})
$, and any term containing an even
number of $\Gamma_{\ul a}$'s and an odd number of $\Gamma_{\ul{\alpha}}$
vanishes.

We can now choose the same gamma matrices used in the Wilson loop case,
eq. (\ref{gammamat}),
and the $\kappa$-fixing can be implemented as in (\ref{spinors8d}):
\bea \tilde\theta^1=\frac1{\sqrt2}\left(\begin{array}{c} \Theta^1 \\
0\end{array}\right)\ ,\qquad
\tilde\theta^2=\frac1{\sqrt2}\left(\begin{array}{c} 0\\ \Theta^2
\end{array}\right)\ , 
\eea    where $\Theta^I$ are again two Euclidean 8-d Majorana-Weyl
spinor of the same chirality.  Then, the
kappa-fixed quadratic fermionic action becomes  \bea\label{actionf1}
S_F&=&-\frac{i}{2\pi}\int d\tau d\sigma \Big[
k|\cos\sigma|\big(\Theta^{1T}\partial_+\Theta^1
+\Theta^{2T}\partial_-\Theta^2\big)+\cr &&\quad\quad\quad\quad-\frac
k4{\rm sign}(\cos\sigma)\sin\sigma\big( \Theta^{1T}\Theta^1
-\Theta^{2T}\Theta^2\big)+\cr  &&\quad\quad\quad\quad-\frac38
k^2\cos^2\sigma\big(\Theta^{1T}\tilde\gamma\Theta^2
-\Theta^{2T}\tilde\gamma\Theta^1\big)\Big]\ ,  \eea  where
$\tilde\gamma=\gamma_{\ul{6789}}$.

Performing the Weyl rescaling of the fermions $\Theta^I\rightarrow
\Theta^I/\sqrt{k|\cos\sigma|}$, we obtain
\bea
\label{actionf}
S_F&=&-\frac{i}{2\pi}\int d\tau d\sigma
\Big[\big(\Theta^{1T}\partial_+\Theta^1
+\Theta^{2T}\partial_-\Theta^2\big) -\frac38
k|\cos\sigma|\big(\Theta^{1T}\tilde\gamma\Theta^2
-\Theta^{2T}\tilde\gamma\Theta^1\big)\Big]\ .\nonumber  \eea  
Thus we find that the eight fermionic fluctuations  have the same
$\sigma$-dependent effective mass  \be
\label{reggeff} m_F=
\frac{3}{4}k\cos\sigma\ \equiv k\,\ell_F \cos\sigma\ .
\ee 
The equations of motion become
\bea\label{reggeqs}
\partial_+\Theta^1-\frac{3k}{8}|\cos\sigma|\tilde\gamma\Theta^2=0\ ,\cr
\partial_-\Theta^2+\frac{3k}{8}|\cos\sigma|\tilde\gamma\Theta^1=0\ ,
\eea
which can be squared to yield the following equations on a flat world-sheet
\bea\label{eqfdiag}
\Big[\partial^\alpha\partial_\alpha+\tan\sigma(\partial_\tau+\partial_\sigma)-\frac{9k^2}{16}\cos^2\sigma\big]\Theta^1=0\ ,\cr
\Big[\partial^\alpha\partial_\alpha-\tan\sigma(\partial_\tau-\partial_\sigma)-\frac{9k^2}{16}\cos^2\sigma\big]\Theta^2=0\ .
\eea
Note that this squared equations are not valid at the degenerate points $\sigma=\pm \frac\pi2$, 
since the equations  have been derived by writing one of the
$\Theta^I$ in term of the other 
from the equations (\ref{reggeqs}), 
and this operation is clearly degenerate at the turning points. Nevertheless, this 
equation must be valid for $\sigma\neq\pm \frac\pi2$ and shows how the parameter $k$ enters the diagonalized 
equations with its square $k^2$, as in the bosonic case. This will be useful in the following sections.

Let us write the squared equations in a non-degenerate
non-diagonalized form by adopting 
the 2-d Majorana spinor formalism.
The equations of motion in the flat world-sheet gauge become
\bea
\big[\tau^\alpha\partial_\alpha-k\ell_F\cos\sigma\,\tilde\gamma\big]\psi=0\ ,
\eea
where the index $i$ of $\psi^i$ is understood. These equations can be squared into
\bea
\big[\partial_\alpha\partial^\alpha - k\ell_F \partial_\sigma|\cos\sigma|\,\tilde\gamma\,\tau_{\ul 1}- 
k^2\ell_F^2\cos^2\sigma\big]\psi=0\ .
\eea 

Unfortunately, the nontrivial $\sigma$-dependence
of the bosonic and fermionic masses does not allow a simple solution of the spectral problem.
In the other cases considered in this paper it is possible to compute the characteristic frequencies 
and to show explicitly how the one-loop energy is finite and then the theory is consistent. 
In this case we will be able to attack the spectral problem by
considering the limits of large and small bosonic and fermionic
masses: 
$k\ell_{B,\, F} \ll 1$ and $k\ell_{B, \, F} \gg 1$. We will do so by implementing perturbative 
techniques usual in quantum mechanics.
The problem of the finiteness of the theory is instead analyzed in appendix \ref{appweyl}.

As usual, we will solve the spectral 
problem by looking for eigenfunctions with a $\tau$ dependence of the form $\sim e^{i\omega \tau}$ 
and solving the resulting spectral equation for the characteristic frequency $\omega$.
For the two massive bosons, we obtain the spectral equation
\bea\label{spectralb}
\big[-\frac{d^2}{d\sigma^2}+k^2\ell_B^2\cos^2\sigma\big]\,y =\omega_B^2\,y\ ,
\eea 
while for the eight fermions the spectral equations read
\bea\label{spectralf}
\big[-\frac{d^2}{d\sigma^2}+k\ell_F \partial_\sigma|\cos\sigma|\,\tilde\gamma\,\tau_{\ul 1}+
  k^2\ell_F^2\cos^2\sigma\big]\,\psi=\omega_F^2\,\psi\ ,
\eea   
Since in the diagonalized equations (\ref{eqfdiag}), $k\ell_{F}$ enters the 
equations of motion with its square, we can  expect that $\omega$ admits an expansion in powers of $(k\ell_{B,\, F})^2$ 
or $1/(k\ell_{B,\, F})^2$ in the $k\ell_{B,\, F} \ll 1$ and $k\ell_{B,\, F} \gg 1$ limits respectively.

\subsection{Small bosonic and fermionic masses}\label{kmin}

Let us first consider the bosons. The equation (\ref{spectralb}) could be solved in terms of Mathieu functions, 
determining the characteristic frequencies in an expansion in $k^2$, as done in \cite{regge}. 
Since we are interested in the first order correction, it is more direct to consider the term 
\bea
V_B=k^2 \ell_B^2 \cos^2\sigma
\eea
as a perturbation to the equation and to use the standard techniques of perturbation theory of nonrelativistic
quantum mechanics. The most useful choice for the eigenfunctions of the unperturbed equation is given by the 
following complex basis
\bea
\langle\sigma|0\rangle=\frac{1}{\sqrt{2\pi}}\ ,\qquad \langle\sigma|n,\pm\rangle=\frac{1}{\sqrt{2\pi}}e^{\pm in\sigma}\ .
\eea
It follows immediately that the squared frequencies are given by
\be
\omega^2_{B(n,\pm)}=n^2+\langle n,\pm|V_B|n,\pm\rangle +{\cal O}(k^4)= n^2+\frac{k^2\ell_B^2}{2} +{\cal O}(k^4) \ ,
\ee
and then
\bea
\omega_{B(n,\pm)}&=&\sqrt{n^2+\frac{k^2\ell_B^2}{2}+{\cal O}(k^4)}\ .
\eea

Turning now to the fermionic case, the perturbation term in (\ref{spectralf}) is the sum of  two terms, 
$V_F=V_{F1}+V_{F2}$, where  
\bea
V_{F1}=k\ell_F \partial_\sigma|\cos\sigma|\,\tilde\gamma\,\tau_{\ul 1}\ ,\qquad  V_{F2}=k^2\ell_F^2\cos^2\sigma\ .
\eea
Since we are interested in the corrections of order $(k\ell_F)^2$, we must compute the first and  second order corrections 
given by $V_{F1}$ and the first order corrections given by $V_{F2}$. Analogously to the bosonic case, $V_{F2}$ 
gives the following correction of order $(k\ell_F)^2$ 
\bea
\delta_{2}\omega^2_{F(n,\pm)}=\frac{k^2\ell_F^2}{2}\ .
\eea
On the other hand, it is easy to see that  $V_{F1}$ does not give any correction at order $k$, since 
$\langle n,\pm|V_{F1}|n,\pm\rangle =0$, while it gives a nontrivial correction  of order $k^2$. 
By using the result
\bea
\left| \int_0^{2\pi}\frac{d\sigma}{2\pi}\partial_\sigma|\cos\sigma|e^{ir\sigma}\right|^2=
\left\{ \begin{array}{l} 0 \quad {\rm if}\quad r=2q+1\ , \\
                         \frac{16}{\pi^2}\frac{q^2}{(4q^2-1)^2}\quad {\rm if}\quad r=2q\  , \\
\end{array}\right.
\eea
where $r$ and $q$ are integer numbers, it is straightforward to obtain the following corrections by using 
standard perturbation theory at second order
\bea
\delta_{1}\omega^2_{F(0)}&=& -\frac{8k^2\ell_F^2}{\pi^2}\sum_{q\geq 1}\frac{1}{(4q^2-1)^2}\equiv -c k^2\ell_F^2\ ,\cr
\delta_{1}\omega^2_{F(n,\pm)}&=& \frac{-4k^2\ell_F^2}{\pi^2}\sum_{q\neq -n}\frac{q}{(n+q)(4q^2-1)^2}\equiv -c_n k^2\ell_F^2\ .
\eea
The series appearing above are convergent and thus we can give the explicit expression for the constants previously introduced
\bea
c&=&{(\pi^2-8)\over2\pi^2}\approx 0.095\ , \cr
c_n&=& {8+96n^2-(4n^2-1)^2\pi^2\over2\pi^2 (4n^2-1)^3}\ .
\eea
Then the zero mode frequency is given by
\bea
\omega_{F(0)}&=&\frac{k\ell_F}{\sqrt2}\sqrt{1-2c}\ .
\eea
On the other hand the non-zero modes get a correction of the form
\bea
\omega_{F(n,\pm)}&=&\sqrt{n^2+\frac{k^2\ell_F^2}{2}-c_n k^2\ell_F^2 +{\cal O}(k^4)}\ ,
\eea
and then the one-loop correction to the spacetime energy, after re-inserting the dimensional factor, is given by
\bea
E_1&=&\frac{m_0}{2k}\Big\{6\times2\sum_{n\geq 1}n+2\times\Big[\frac{k\ell_B}{\sqrt{2}}+ 
2\sum_{n\geq 1}\sqrt{n^2+\frac{k^2\ell_B^2}{2}+{\cal O}(k^4)}\Big]+\cr
&&-8\times\Big[ \frac{k\ell_F\sqrt{1-2c}}{\sqrt{2}} +2\sum_{n\geq 1} \sqrt{n^2+\frac{k^2\ell_F^2}{2}-c_n k^2\ell_F^2+{\cal O}(k^4)}\Big]\big\}\ .
\eea
By expanding in powers of $(k\ell_B)^2$ and $(k\ell_F)^2$, using  the
mass matching condition $2\,\ell_B^2=8\,\ell_F^2$ 
and the fact that $c_n \sim 1/n^2$ for large $n$, it is easy to see that the sum is indeed finite 
and that the zero modes contribute to a correction at order $1$ while the other modes at order $k$, 
leading to
\bea
E_1= z_0+k z_1+{\cal O}(k^2)\ ,
\eea
where $z_0$ and $z_1$ are given by
\bea
z_0=\frac{m_0}{\sqrt{2}}[\ell_B-4\ell_F\sqrt{1-2c}]= \frac{3m_0}{\sqrt{2}}[{1\over2}-\sqrt{1-2c}]\approx -0.85\,m_0\ ,
\eea
and
\be
z_1=\frac{4m_0}{k^2}\sum_{n\geq 1}{c_n k^2\ell_F^2\over n}= {2\ell_F^2m_0\over\pi^2}[-24+\pi^2+16\log2-\pi^2\log4+14\zeta(3)]\approx 0.012\,m_0\ .
\ee
The limit of small $m_B$ and $m_F$ requires that $\lambda\gg J$. In particular, 
by considering $J$ finite (and then $E_0$ of order $\sqrt\lambda$) and neglecting all terms 
of order $1/\lambda^2$, the corrected 
Regge trajectory takes the following form 
\bea
\label{rt}
J=\alpha^\prime_{adj}\Big(1-\frac{6z_1}{\lambda\,m_0}\Big)\left[E^2 -2E\,z_0+ z_0^2\right] .
\eea
where $\alpha^\prime_{adj}=3/2\lambda m_0^2$. Thus we see that the
effect of the $z_1$ term is to shift the effective slope of the Regge
trajectory, while $z_0^2$ gives a positive intercept. Notice that the
``renormalization'' effect on the effective slope occurs in the form
$1-a^2/\lambda$: thus the effective ``adjoint tension'' is rescaled as
$T_{adj}^{(ren)}\sim {1\over 2\pi \alpha'_{adj}}(1+a^2/\lambda)$,
i.e. with an opposite sign w.r.t. the other cases examined in the
paper. Finally notice that we are still restricting ourselves to a
large $J$ regime in order to thrust the semi-classical approximation.

\subsection{Large bosonic and fermionic masses}\label{kmag}

In order to present a unified treatment of the bosonic and fermionic
contributions we find it convenient to think about the large-mass
limit as a large-$k$ limit. Note that  equivalently we have 
a large $J/\lambda$ limit  for $k\gg 1$.
To solve the spectral problem we first define $\omega\equiv k\alpha$.

Let us start with the bosonic case. The spectral equation (\ref{spectralb}) takes the form
\bea\label{spectralb2}
\big[-\frac{1}{k^2}\frac{d^2}{d\sigma^2}+\ell_B^2\cos^2\sigma\big]\,y =\alpha_B^2\,y\ .
\eea 
This can be seen as a one-dimensional stationary Schr\"odinger equation, where $1/k$ plays the role of $\hbar$, 
$\alpha_B^2$ is the energy and $V_B(\sigma)\equiv \ell_B^2\cos^2\sigma$ the potential. 
Then the $k\gg 1$ regime corresponds to the quasi-classical regime $\hbar\ll 1$, where the WKB 
approach is reliable. It consists of making the formal substitution
$y(\sigma)=e^{ik\chi(\sigma)}$, to obtain the equation
\bea
(\chi^\prime)^2-\frac{i}{k}\chi^{\prime\prime}=\alpha^2-\ell_B^2\cos^2\sigma\ .
\eea 
Since we are interested only in the leading order result in the $k\gg 1$ limit, we can simply drop the 
$1/k$ term. Then the above equation can be easily integrated 
\bea
\chi(\sigma)=\pm\int_0^\sigma d\tilde\sigma\sqrt{\alpha^2-\ell_B^2\cos^2\tilde\sigma}\ .
\eea  
We have then an oscillating wave function in the region where $\alpha^2> \ell_B^2\cos^2\sigma$ while a 
rapidly decreasing exponential wave function where $\alpha^2< \ell_B^2\cos^2\sigma$. 
In particular, at the turning points $\bar\sigma$ where $\ell_B^2\cos^2\bar\sigma=\alpha^2$ the WKB 
approximation looses its reliability.

If $\alpha^2> \ell_B^2$ we can use the WKB approximation for any point $\sigma\in[0,2\pi]$. For a given $\alpha^2$
two independent symmetric and antisymmetric eigenfunctions are then given by
\bea
y\sim \cos\Big\{k\int_0^\sigma d\tilde\sigma\sqrt{\alpha^2-\ell_B^2\cos^2\tilde\sigma}\Big\}\ ,\qquad
y\sim \sin\Big\{k\int_0^\sigma d\tilde\sigma\sqrt{\alpha^2-\ell_B^2\cos^2\tilde\sigma}\Big\}\ .
\eea
By imposing the periodicity condition 
\bea
k\int_0^{2\pi} d\tilde\sigma\sqrt{\alpha^2-\ell_B^2\cos^2\tilde\sigma}=2\pi n\ ,
\eea
we get
\bea\label{specell}
E(\zeta_B^2)=\frac{\pi}{2\ell_B}\zeta_B x\ ,
\eea
where we have introduced $\zeta_B\equiv \ell_B/\alpha\in(0,1]$, $x=n/k$ and $E(\zeta^2)$ is the complete elliptic integral of the
second kind
\bea
E(\zeta^2)=\int_0^{\pi/2} d\sigma\sqrt{1-\zeta^2\sin^2\sigma}\ ,\qquad \zeta^2< 1\ .
\eea

When $\zeta_B\geq 1$, in the $k\rightarrow \infty$ limit the corresponding quantum mechanical particle lives in one of the 
two potential wells, with subleading tunnel effect. Choosing the domain $\sigma\in[-\pi,\pi]$, we can 
then consider an eigenfunction $y_0(\sigma)$ 
concentrated in the right well and construct the two eigenfunctions (symmetric and antisymmetric) 
\bea
y_\pm(\sigma)\sim y_0(\sigma)\pm y_0(-\sigma)\ .
\eea 
These correspond to the same eigenvalue, up to exponentially suppressed terms.
If $\sigma_{min,max}\in [0,\pi]$ are the turning points ($\sigma_{max}=\pi-\sigma_{min}$), 
the Bohr-Sommerfeld quantization condition reads 
\bea
k\int_{\sigma_{min}}^{\sigma_{max}} d\sigma\sqrt{\alpha^2-\ell_B^2\cos^2\sigma}=\pi (n+\frac12)\ .
\eea 
In the large $n,k$ limit with $x=n/k$ fixed, we can rewrite this relation as 
\be
F(\zeta^2_B)=\frac{\pi}{2\ell_B}\zeta_B x\ , \qquad 2F(\zeta^2)\equiv\int_{\sigma_{min}}^{\sigma_{max}} d\sigma\sqrt{1-\zeta^2\cos^2\sigma}\ ,\qquad \zeta^2\geq 1\ .
\label{nea}
\ee
This expression is continuously connected with (\ref{specell}), both approaching the same expression when $\zeta_B\rightarrow1$. In the following we will recall $E$ the function covering the whole interval $\zeta\in(0,\infty)$. Let us notice that, from (\ref{nea}) it can be deduced that when $\zeta\rightarrow\infty$, then $x\rightarrow0$. 

Let us now turn to the fermionic case. After the substitution $\omega_F=k\alpha_F$ the  equation (\ref{spectralf})
becomes
\bea\label{spectralf2}
\big[-\frac1{k^2}\frac{d^2}{d\sigma^2}+\frac1k\ell_F \partial_\sigma|\cos\sigma|\,\tilde\gamma\,\tau_{\ul 1}+
 \ell_F^2\cos^2\sigma\big]\,\psi=\alpha_B^2\,\psi\ .
\eea  
It is now easy to see that by making the formal substitution $\psi=e^{ik\chi}$ and keeping again only the 
leading order terms, we obtain the very same spectral equation of the bosonic case,
\bea
(\chi^\prime)^2=\alpha^2-\ell_F^2\cos^2\sigma\ ,
\eea 
with $\ell_F$ instead of $\ell_B$.
We can then use all the results obtained for the bosonic case by simply substituting $\ell_B$ and $\zeta_B$ with 
$\ell_F$ and  $\zeta_F$ respectively. 

The one-loop correction to the energy, up to terms of order $1/k$, can be expressed as
\bea
E_1&=&\frac{k m_0}{2}\lim_{\Lambda\rightarrow \infty}\Big\{ 6\times 2 \int_0^\Lambda dx\, x +2\times 2\int_0^\Lambda dx\, \alpha_B(x)
- 8\times 2\int_0^\Lambda dx\, \alpha_F(x)\Big\}\ . 
\eea
Using (\ref{specell}) and the corresponding fermionic equation it is possible to change the variable of integration from $x$
to $\zeta_{B,F}$, obtaining
\bea
E_1&=&\frac{k m_0}{2}\lim_{\Lambda\rightarrow \infty}\Big\{6\Lambda^2+
2\times \frac{4\ell_B^2}{\pi}\Big( \Big[\frac{E(\zeta^2)}{\zeta^2}\Big]_{+\infty}^{\ell_B/\Lambda}+
\int_{+\infty}^{\ell_B/\Lambda}d\zeta\frac{E(\zeta^2)}{\zeta^3} \Big)+\cr
&&-8\times \frac{4\ell_F^2}{\pi}\Big( \Big[\frac{E(\zeta^2)}{\zeta^2}\Big]_{+\infty}^{\ell_F/\Lambda}+
\int_{+\infty}^{\ell_F/\Lambda}d\zeta\frac{E(\zeta^2)}{\zeta^3} \Big)\Big\}=\cr
&=&\frac{k m_0}{2}\lim_{\Lambda\rightarrow \infty}\Big\{6\Lambda^2+
2\times \frac{4\ell_B^2}{\pi}\Big( \Big[\frac{E(\zeta^2)}{\zeta^2}\Big]_{\ell_F/\Lambda}^{\ell_B/\Lambda}+
\int_{\ell_F/\Lambda}^{\ell_B/\Lambda}d\zeta\frac{E(\zeta^2)}{\zeta^3} \Big)\Big\}  \ ,
\eea
where in the last step the mass matching condition $2\ell_B^2=8\ell_F^2$ has been used\footnote{Notice that, due to the mass matching condition, the region $\zeta\in[1,\infty)$ does not contribute to $E_1$ and so the details of the function $F$ defined as in (\ref{nea}) are not necessary (we only need to know its behavior for $\zeta\rightarrow\infty$).}.

By expanding $E(\zeta^2)$ for small $\zeta^2$, it is possible to show that the quadratic and logarithmic divergences
drop out and we are left with the following one-loop finite correction to the energy
\be
E_1=-\frac{k m_0}{4} \Big[2\times \frac{\ell_B^2}{2}\log \frac{\ell_B^2}{2}- 8\times \frac{\ell_F^2}{2}\log \frac{\ell_F^2}{2}\Big]= -m_0(\frac{9}{8}\log 2) \, k\ .
\ee    
Let us observe that this correction is negative as in the other cases considered in this paper and that it has 
the same universal form, with the $\sigma$ dependent squared masses $m^2(\sigma)=k^2\ell^2\cos^2\sigma$ 
 substituted by their mean value $m^2=k^2\ell^2/2$.  
The effect of the one-loop correction in Regge trajectory translates in a ``renormalization'' (actually a rescaling) of the effective string tension
\bea\label{renregge}
E^2=4\pi T_{QCD}^{(ren)}(\lambda)J\ ,
\eea
with
\bea\label{trenreg}
T_{QCD}^{(ren)}(\lambda)=\Big(1-\frac{27\log 2}{8\lambda}\Big)^2T_{QCD}\ .
\eea 
This is equal to the renormalized effective string tension (\ref{eloopfin}) obtained  
from the one-loop correction to the  Wilson loop energy. 
The relation $T_{adj}=2T_{QCD}$ is then preserved by quantum corrections in this regime of parameters.

Notice that in the above calculations we have suppressed the subleading orders in the large $k$ limit. Among these there is a (negative) $1/k$ term coming from the six massless modes, shifting the value of $E_1$. 
In particular, this term would produce, as in the $k\ll 1$ regime, a non-linearity of the Regge trajectory and a positive intercept, which are thus fairly general outcomes of the sigma-model calculations.  
We did not write this term explicitly since it is possible that we are neglecting similar terms while using the WKB approximation.

\section{Strings spinning in internal directions: Stringy hadrons}\label{int}  

In this section we concentrate on a class of string solutions related
to sectors of the gauge theory which do not have a direct contact with
pure YM. In fact what we are going to consider are string states with
large spins along the transverse $S^4$ of our background.  These
correspond to multi-charged gauge theory hadrons whose constituents are
the adjoint massive fields \cite{gpss,alfra,son,gfacl}.

The bosonic action (\ref{bosac}), with $X^i={\rm const.},\,
i=1,...,3$, reduces to a case widely studied in literature, which  can
be connected to the $n=4$ Neumann integrable system, as already done
in \cite{tomilu} following the general approach adopted for the search
of string solutions on $AdS_5\times S^5$ (see \cite{tsey} and
references therein). At the classical level, in fact, this is exactly
analogous to the case considered in \cite{frolov5,russo,tsey} and
reviewed in \cite{tomilu}. In the latter paper some solutions were studied,
corresponding to circular and folded strings rotating on the internal
$S^3$ of the IR region of the (softly broken) Maldacena-N\`u\~{n}ez (MN)
background. These solitonic stringy states were interpreted as
corresponding to multi-charged hadronic states in  the dual gauge
theory. We can readily adapt the results of \cite{tomilu} to the
present discussion.
  
First of all, it is useful to introduced the following
parameterization of the $S^3\subset S^4$: \bea\label{curved}   &&
Z^1=\sin\chi\cos\phi_-\ ,\qquad Z^2=\sin\chi\sin\phi_-\ ,\cr   &&
Z^3=\cos\chi\cos\phi_+\ ,\qquad Z^4=\cos\chi\sin\phi_+\ , \eea   in
which the $S^4$ metric becomes \bea
d\Omega^2_4=d\psi^2+\cos^2\psi(d\chi^2+\sin^2\chi d\phi_-^2+\cos^2\chi
d\phi_+^2)\ .  \eea   In the case of the folded strings, the solution
reads ($\psi$ is always kept constant at zero value) \be
\theta_2(\sigma)=\theta_2(\sigma+2\pi)\ , \qquad X^0=k\tau\ , \qquad
\phi_+=\nu\tau\ ,\qquad \phi_-=\omega\tau\ ,   \ee   where
$\theta_2=2\chi$.   The Virasoro  constraint implies that
\be\label{constraint2}   \theta_2'^2-2(\omega^2-\nu^2)\cos{\theta_2}=4
k^2  - 2(\omega^2+\nu^2)\ .   \ee   The conserved charges are
\bea\label{chs}   E &=& \xi k\ ,\cr   J_+ &=&
\frac{\xi\nu}{4\pi}\int_0^{2\pi}d\sigma(1+\cos\theta_2(\sigma))\ ,\cr
J_- &=&
\frac{\xi\omega}{4\pi}\int_0^{2\pi}d\sigma(1-\cos\theta_2(\sigma))\ ,
\eea   from which it follows that  \be   E=\frac{k}{\nu}J_+
+\frac{k}{\omega}J_-\ .  \ee   There are various special cases. For
$\omega^2=\nu^2$ the solution describes circular strings and is given
by  $\theta_2 = \pm 2\sigma\sqrt{k^2-\omega^2}+{\rm const}$. This is
in general an extended solution, except for $k^2 =\omega^2$ where it
describes a point-like string. The periodicity condition implies the
energy/charge relation $E=\sqrt{m^2\xi^2+J^2}\ ,$   where $m$ is the
number of windings and $J=J_+ + J_-$ is the sum of the two charges.
  
In the general case $\omega^2\neq \nu^2$, we can obtain the
energy/charge relation  in the short and long string limits.   For
short strings  \be   E\sim J + \frac{\xi^2J_-}{2J^2}\ ,   \ee   which
is the analogous of the BMN expansion.   In the long string limit, the
relation is  \be   E\sim J + \frac{2\xi^2}{\pi^2J_+}\ .   \ee

The homogeneous circular string solutions have instead the form
\bea\label{hom}   Z^1+iZ^2=a_1e^{iw_1\tau+im_1\sigma}\ ,\quad\quad
Z^3+iZ^4=a_2e^{iw_2\tau+im_2\sigma}\ ,\quad\quad X^0=k\tau\ ,    \eea
with $w_i,\ m_i,\ i=1,2$, satisfying the following relations
\bea\label{relat}   & w_i^2=m_i^2-\Lambda\ ,& a_1^2+a_2^2=1\ ,\cr   &
k^2=2(a_1^2w_1^2+a_2^2w_2^2)+\Lambda\ ,&a_1^2w_1m_1+a_2^2w_2m_2=0\ .
\eea   In the coordinates (\ref{curved}), the solutions (\ref{hom})
take the form   \bea\label{solcurv}   X^0=k\tau\ ,\qquad \sin\chi=a_1\
,\qquad \phi_-=w_1\tau+m_1\sigma\ ,\qquad \phi_+=w_2\tau+m_2\sigma\ .
\eea   The charges are $J_-=\xi a_1^2w_1\ , J_+=\xi a_2^2w_2$.  In the
large $J$ limit one observes that  \bea
\label{energy-spincirc}  
E=J\Big[1+\frac{\xi^2}{2J^2}\Big(m_1^2\frac{J_{-}}{J}+m_2^2\frac{J_+}{J}\Big)+\ldots\Big]\
.   \eea

All we have said up to now concerns the classical solutions.  One can
go further and try to calculate the leading one-loop sigma-model
corrections to the relations above.   This amounts to  a calculation of
the vacuum energy of the string theory expanded to the first
non-trivial order around the classical solution.  While for the folded
strings there are some subtleties in the quantization procedure, due
to the non-trivial world-sheet metric (see for example
\cite{frolov1}), for the circular solutions the calculation is
straightforward.   Since the coordinates $\phi_-$ and $\phi_+$ never
appear in the components of the background fields, they enter the
string action only through the pull-back of background fields. This
means that, in the expansion of the string action around the solution
(\ref{solcurv}), all the coefficients are  constants depending only
from $k,w_i$ and $m_i$. The same holds for all the conserved charges.
  
Even in the simple example of the circular solutions on the
fixed-radius-$S^5$ in $AdS_5 \times S^5$, the calculation of the vacuum energy is 
very difficult, and only its leading contribution can be
computed explicitly, in the limit of very large $k$
\cite{frolov3,frolovnew}. The same holds for the more involved MN solution
\cite{tomilu}, and we expect the very same behavior in our case.  But,
as far as the large $k$ limit (for fixed $m_i$) is concerned, we can
make the following observation.   From (\ref{relat}) and
(\ref{solcurv}) it follows that in this limit we can neglect at the
leading order the contributions coming from the $m_i$.   As such, the
one-loop expansion is equivalent, in the limit, to the expansion
around a collapsed, {\it point-like} string rotating on $S^4$ with an
angular velocity $\nu=\sqrt{a_1^2w_1^2+a_2^2w_2^2}$.  This in turn is
the same as studying the exact (in $\alpha'$) string theory on the
Penrose limit of the original background.   Then, in the large $k$
limit the leading one-loop quantum correction to the energy of the
circular configurations is identical to that obtained in the Penrose
limit, i.e. on the corresponding pp-wave.    This has been explicitly
verified in $AdS_5 \times S^5$ \cite{frolov3}, where the vacuum energy
is vanishing at this order    because of the linearly realized
supersymmetries in the Penrose limit, and for the MN solution in
\cite{tomilu}.  We are then going to study the Penrose limit theory in
detail in the following section.  But let us just quote here for
completeness the result for the energy-charge relation of the circular
strings (we re-insert the dimensional factor and use $\xi=\lambda/3$)
\be\label{etot}
E=m_0J\Big[1+\frac{\lambda^2}{18J^2}\Big(m_1^2\frac{J_{-}}{J}+m_2^2\frac{J_+}{J}\Big)+\ldots\Big]+m_0J\Big[\frac{45}{8\lambda}\log{\frac34}+\ldots\Big]\
.    \ee   The sigma-model result provides a strong $\lambda$ coupling
renormalization of $m_0J$.  While being a subleading term for large
$\lambda$, it forbids to extrapolate the classical result
(\ref{energy-spincirc}) to the small coupling regime as for the
spinning strings in $AdS_5 \times S^5$, and so to compare this  result
with a field theory calculation.  We will comment further about this
result and about the field theory duals to the spinning strings in
section \ref{hadr}.

\subsection{The plane wave theory}\label{ppw}  
  
The semiclassical quantization around the point-like string spinning
at the speed of light around the large equator of the four-sphere is
equivalent to the study of the string theory on the plane-wave
background, obtained as a Penrose limit of the original metric
(\ref{defns}).  
We present the Penrose limit in appendix \ref{apppenr} and just quote the result: 
\bea   ds^2&=&-4dx^+dx^- - m_0^2[v_3^2+v{\bar v} +
{3\over4}u{\bar u}]dx^+dx^+ + dx^idx^i + dud{\bar u} + dvd{\bar v} +
dv_3^2\ ,
\label{pmetf}\cr  
e^{\Phi_0}F_4 &=& {3i\over2} m_0 dx^+\wedge dv_3\wedge
dv\wedge d{\bar v}\ .   
\eea
The dilaton becomes a constant $e^{\Phi_0}\approx g_sN^{1/2}$.
In the plane wave metric above, the three coordinates $x^i$ come from the three flat special directions,
the three $v$ modes from the four-sphere (the fourth mode of the sphere combines with the time
variable into the light-come coordinates $x^+,\ x^-$), and the $u$ modes from the radius and circle direction of the original background. 
The $x^i, v$ fields parameterize the
so-called universal sector. The IIA Hamiltonian can be read as the
one describing the non relativistic motion in flat 3-d space ($x^i$)
of gauge theory hadrons made of adjoint massive constituents, along the lines of \cite{gpss}.
The conserved light-cone Hamiltonian $H$ and momentum $P^+$ read as usual   \be
\label{d2fd2h}  
H =i\partial_+    = E -m_0J\ ,\qquad
P^+ ={i\over2}\partial_{-}= {m_0\over
\lambda}J\ .   \ee

Let us now study the light-cone gauge IIA GS string on the above
background. There are three massless bosonic fields corresponding to
the three spatial directions $x^i, i=1,2,3$. The remaining five
massive fields have frequencies   \be   \omega^{v,\bar{v}}_n =
\sqrt{n^2 + m^2}\ , \qquad \omega^{v_3}_n= \sqrt{n^2+m^2}\ ,\qquad
\omega^{u,\bar{u}}_n= \sqrt{n^2 +{3\over4}m^2}\ ,   \ee where
$m=m_0\alpha'p^+$.
  
Concerning the fermionic fields, their equations of motion read   \bea
(\partial_{\tau} + \partial_{\sigma})\theta^1 &=& -
{\alpha'p^+\over{(4) 3!}}e^{\Phi}F_{+ijk}\Gamma^{ijk}\theta^2\ ,
\nonumber \\   (\partial_{\tau} - \partial_{\sigma})\theta^2 &=& -
{\alpha'p^+\over{(4) 3!}}e^{\Phi}F_{+ijk}\Gamma^{ijk}\theta^1\ ,
\eea   from which we get that the fermionic fields are all massive,
and all have frequency   \be   \omega^I_n= \sqrt{n^2 +{9\over16}m^2}\
, \qquad I=1,...,8\ .   \ee   

The light-cone Hamiltonian can be obtained straightforwardly  and can be expressed as a sum of 
a contribution from the momentum and the stringy excitations in the three flat dimensions
and a contribution from the massive ``zero'' modes and excitations of the internal directions 
and the fermionic fields (see \cite{gpss}).  

Since the sum of the squares of the fermionic frequencies exactly matches with the analogous
bosonic sum, our model is finite with nontrivial zero-point energy 
\be \label{zerop}
E_0(m)={m_0\over 2m}\sum_{n=-\infty}^{\infty}\Biggl[ 3n + 3\sqrt{n^2 +
m^2} + 2\sqrt{n^2+{3\over4}m^2} - 8 \sqrt{n^2+ {9\over16}m^2}\Biggr]\
.   \ee   Following the general arguments in
\cite{alfra2,alfra3,gfacl} we see that this is negative and increases
in absolute value as $m$ in the large $m$ limit.   The latter is also
the large $k$ limit of the circular string configuration, so it is
interesting to exhibit its explicit leading order form   \be \label{zerop}
E_0(m)\approx \frac{15m_0m}{8}\log{\frac34}\ .   \ee In the small m
limit, instead, we have \be   E_0(m)\to - {(3-{\sqrt3})\over2}m_0\ .
\ee   We see that the energy/charge relation for the point-like
strings in the large $m$ regime reads\footnote{From
(\ref{d2fd2h}) and (\ref{david}) it follows that
$m=3J/\lambda$.}   \be \label{etotpoint}
E=m_0J(1+\frac{45}{8\lambda}\log{\frac34}+...)\ .  \ee

In section \ref{hadr} we will comment about the field theory duals of
our plane-wave string states.

\section{Concluding remarks on the dual field theory}\label{com} 

Let us summarize the implications of our string theory
considerations for the dual field theory.  A very generic property of
the  the leading order one-loop corrections to the  energies of the various
configurations we have analyzed is that they are always
negative. Their precise meaning can be sketched case by case.

\subsection{Wilson loop}\label{willoop}

Let us first discuss  the string configuration
corresponding to the Wilson loop. Our main result is the
fact that for a worldsheet supersymmetric strings on the confining
Witten background, apart from the L\"uscher $1/L$ term, the leading order one-loop correction to the quark-antiquark
potential, in the large $L$ limit, gives a ``renormalization'' of the YM string tension\footnote{
Interestingly , this ``renormalization'' of the string tension (without L\"uscher term) takes place, at the classical level, 
in MQCD \cite{cobi}.}.  As Gross and Ooguri discussed
in \cite{gross}, this fact is somewhat expected. The gauge theory
under consideration has a UV cutoff given by $m_0\sim
1/R_{\theta}$ and the coupling $g_{YM}$ should be considered as the
bare coupling at scales of order $m_0$. As we have recalled, the
classical string tension in the strong 't Hooft coupling limit is
given by
\bea\label{lim1}
T_{QCD}=\frac1{6\pi}\lambda m_0^2\quad\quad  {\rm for}\quad\quad\lambda=g_{YM}^2N\gg 1\ . 
\eea 
To construct the pure QCD we should consider the limit of very large UV cutoff $m_0$ and the 't Hooft coupling constant should be given by the usual perturbative expression
\bea
\lambda=\frac{b}{\log \left(\frac{m_0}{\Lambda_{QCD}}\right)}\ ,
\eea   
where $b$ is given by the first coefficient of the beta function. Gross and Ooguri suggested that a reasonable behavior of  the tension for small $\lambda$ should be
\bea\label{UVtension}
T_{QCD}=ae^{-\frac{2b}{\lambda}}m_0^2=a\Lambda_{QCD}^2\quad\quad{\rm for}\quad\quad \lambda\ll 1\ ,
\eea
where $a$ is some numerical constant.    
Then, there should be an interpolation function $f(\lambda)$ such that
\bea
T_{QCD}=f(\lambda)m_0^2\ ,
\eea
with the previous asymptotic behaviors (\ref{lim1}) and (\ref{UVtension}). 
In our calculation we have computed $f(\lambda)$ to the second order in the large $\lambda$ limit.

The fact that in the large $L$ limit we have a renormalization of the string tension besides a L\"uscher term depends crucially of three facts: 1) our string is supersymmetric; 2) we consider 
fluctuations around the Witten background dual to YM as opposed to
flat space; 3) we 
evaluate the one-loop correction to the energy ($E_1$) without
separately regularizing the 
bosonic and fermionic contributions, but simply by calculating the 
whole fermion + boson contribution \cite{alfra2}.

Interesting, ignoring any one of these  three conditions results in
the absence of this correction to the tension.  For example,  if we
reject 3) but not 1) and 2) and evaluate the $E_1$ with the
$\Delta$ renormalizations  (i.e. the $\zeta$-renormalization
generalized to allow for  world-sheet masses),  we find  that in the
large $L$ limit the only non vanishing contribution comes from  the
six massless bosonic fields, 
i.e. we are left with only the standard L\"uscher term.  If we
reject point 1) and consider only the bosonic contributions we are
forced to renormalize and again only the L\"uscher term is present.

One way to interpret our result is that it yields falsifiable predictions for the
holographic approach to nonsupersymmetric gauge theories. In
particular, there is a clear distinction about the nature of the ``YM
string''. If the YM string is an intrinsically bosonic object, then we
expect only a L\"uscher term correction. On the other hand, if the
string is intrinsically fermionic\footnote{It has been suggested that
the string dual to YM theory has to include worldsheet supersymmetry
in order to get rid of the open string tachyon \cite{wall}.} then the
most likely scenario for the leading order correction is a renormalization of the
string tension. Future studies of the $q\bar q$ potential should settle this
question. 
 
\subsection{Regge trajectory}\label{regcom}
One of the most distinctive features of the strong interactions is
Regge theory which encompasses the particle spectrum, the forces
between particles and the high energy behavior of scattering
amplitudes. Regge trajectories are one of the most ubiquitous
properties shared by most strongly interacting states. 

The UA8 collaboration \cite{ua8} has presented a description of
the nonlinearities of the Regge trajectory corresponding to  the soft
Pomeron. Based on its quantum numbers, this trajectory 
is believed to be composed of glueball
states, although no direct evidence is available. The best
experimental fit is of the form:
\begin{equation}
J=\alpha(t)=1.10 + 0.25 t + \alpha'' t^2,
\end{equation}
where $\alpha''=0.079\pm 0.012 GeV^{-4}$ and $t$ is the mass squared of the particles.  

The non-linearity and the positive intercept that we find as the result of one-loop corrections
in the framework of IIA string theory are two important qualitative
properties shared 
with this trajectory. It is worth
remarking that this behavior is exhibited by the loop-corrected
trajectories of supersymmetric supergravity backgrounds discussed in
\cite{regge} in the framework of IIB string theory. 
The constancy of this result encourages us to think that
this is indeed a very universal property of confining theories
admitting a holographic description in IIA, IIB, supersymmetric
or nonsupersymmetric cases.  

The string theory non-linearity is in a term
proportional to $\sqrt{t}$ instead of $t^2$.  This different behavior
could be due to the different limits of the $\lambda$ coupling  and
spin $J$ regimes where the two results are relevant. There are
successful 
phenomenological models \cite{brisudova} interpolating between a
squared root and quadratic behavior in $t$. It is also possible
that the $t^2$ non-linearity could be due to mixing with fundamental
matter and decay effects present in the finite $N$ theory and should
disappear in the planar limit. Having its origin in the massive modes
of the worldsheet fluctuations, we expect the $\sqrt{t}$ non-linearity
to be due to mixing with the adjoint matter always present in the
supergravity regime. These massive terms are intrinsically quantum
mechanical and vanish in the string classical result
giving the linear trajectory.

Our evaluation of this non-linear effect is valid for large  $J$ and
much more larger $\lambda$. Then, it is hard to believe that the
details of this result can be extrapolated to the decoupled, pure YM
regime $\lambda\ll 1$. On the other hand in section \ref{kmag} we have
considered  also the limit $J\gg\lambda$, which seems to be more
reliable for a  comparison with the decoupled theory, and in this case
we  have obtained that the leading quantum corrections give a
renormalization  of the effective string tension completely consistent
with the result obtained  from the Wilson loop analysis.
We would like to stress that, in this regime, subleading corrections 
to the result we found could produce again non-linearities in the trajectory,
including $t^2$ ones.
It would be very interesting to calculate these corrections, in order to see if
this picture is actually realized. 

Another qualitative agreement with the experimental data comes from
the comparison with the slopes of the Regge trajectories for mesons
and for the soft  Pomeron. The phenomenological data indicate that the former is
approximately 3.6 times bigger than the latter. The classical string
computation gives instead only a factor of 2, to be traced back to the
difference between open and closed strings, consistently with the
infinite $N$ limit of the Casimir ratio in formula (\ref{fourfour}). According to
the latter, the inclusion in the string calculation of finite $N$
contributions should produce for $N=3$ just a little improvement to a
factor of 2.25.  But the first $\alpha'$ stringy quantum correction
seems to point in the right direction to solve the discrepancy. In
fact, comparing the renormalized fundamental and adjoint string
tensions $T_{QCD}^{(ren)}$ from (\ref{eloopfin}) and $T_{adj}^{(ren)}$
from (\ref{rt}), we read \be\label{compare} {T_{adj}^{(ren)} \over
T_{QCD}^{(ren)}} \sim 2 \Big( 1 + {4.8 \over \lambda} \Big).  \ee The
plus sign in the correction suggests that reducing the value of the
coupling toward the realistic regime the ratio of the tensions
correctly increases. Thus, even if the very large $J$ and $\lambda$
regime where our computation is reliable is not the phenomenological
one, we take formula (\ref{compare}) as an indication that the
inclusion of higher stringy corrections points toward a better
agreement of the string models with the real world\footnote{Note that
in this argument we used the $k \ll 1$ (i.e. $J\ll \lambda$) result.
In the opposite limit of $k \gg 1$ we find no $\lambda$ correction to
the  ratio of the tensions.  This suggests that for generic values of
$k$ the correction should be present, although with a smaller
magnitude with respect to the one in (\ref{compare}).  The main result
is thus the same.}.
As a final caveat, we should stress that in (\ref{compare}) we used the mesonic
string tension calculated via the Wilson loop. It would be very
interesting to check if this value, as expected, is exactly the same one
that appears in the open string calculation of the mesonic Regge
trajectories.

\subsection{Stringy hadrons}\label{hadr}

In this section we comment on the field theory states dual to the
rotating strings of sections \ref{int}.  The general matching between
(some of) the string massless excitations and the field theory
operators, first analyzed in \cite{gpss}, is discussed in
\cite{gfacl}.  Let us review the basic features.

The field theory object dual to our string ground state must be made
out of some massive fields charged under the global $SO(5)$ flavor
group.  As such, the gauge degrees of freedom are ruled out.  The
natural building blocks for describing the field theory states are the
five adjoint scalars $\Phi_i,\ i=1,...,5,$ living on the D4
world-volume before wrapping the supersymmetry breaking cycle.  After
wrapping, they get (the same, since $SO(5)$ is preserved) mass due to
loop corrections. The string prediction is that this mass, in the
strong coupling regime and in the hadron bound state, gets
renormalized to $m_0$.  The scalars transform in the vector
representation of $SO(5)$, so, with respect to the two $U(1)$'s
subgroups, corresponding to the $J_+$ and $J_-$ charges of the
spinning strings (in sections \ref{ppw} $J_+$ was called simply $J$),
we can arrange a complex linear combination, call it $Z$, with charges
(1,0) respectively, a second complex combination $W$ with charges
(0,1), and we are left with an uncharged field $\Phi$.

From these data we can immediately conclude that the ground state of
the string on the plane-wave of section \ref{ppw} is dual to a  hadron
created by acting on the field theory vacuum as
$Tr[Z^{J_+}]|0\rangle_{FT}$. More precisely, in the absence of
state/operator correspondence we can simply identify the hadron as the
lowest energy state created by the action of the given operator on the
gauge theory vacuum. The latter has in fact the right quantum numbers:
its mass is $J_+$ times $m_0$ and its charge is precisely $J_+$, so,
according to the energy formula (\ref{d2fd2h}), its dual string state
has light-cone Hamiltonian $H=E-m_0J_+=0$ and is, therefore, the
perfect candidate for the string ground state  (\ref{etotpoint}).

Let us propose an identification of some of the string theory zero
modes  with gauge theory states. If we look at the modes coming from
the $S^4$ of the original Witten background, we see that the $v,\
\bar{v}$ coordinates of the plane-wave (\ref{pmetf}) are the ones
charged under $J_-$, while $v_3$ is not charged under it, but the
three of them are not charged under $J_+$.  Their zero modes have mass
$m_0$, so their action on the string theoretic ground state gives a
state with energy $H=m_0$.  In field theory we can easily identify
them with the fields $W,\ \bar{W}$ and $\Phi$, respectively: all of
them have mass $m_0$ and zero $J_+$ charge, so their insertion in the
hadron $Tr[Z^{J_+}]|0\rangle_{FT}\ $ gives a total contribution to the
Hamiltonian as $H=m_0(J_++1)-m_0J_+=m_0$.

It was also argued in \cite{gpss} that the three massless string modes
$x_i$ should describe the non relativistic motion (and excitations) of
the hadron in the three special direction in field theory.  We refer
the interested reader to that paper for a discussion of this point.
For what concerns the $u,\ \bar{u}$ modes in (\ref{pmetf}), their
identification has always been very difficult
\cite{gpss,alfra,son,gfacl}, and we do not have any definite proposal.

Let us come to the multi-charged string solitons of section \ref{int}.
The ingredients in field theory are the same as before, the only
difference is that the string ground state  should be identified with
a hadron carrying two very large charges $J_+,\ J_-$. Unfortunately,
this is the only identification we will make given the absence of the
spectrum of excitation.  Identifying the ground state  is an easy task
since  a state created by $J_+$ fields $Z$ and $J_-$ fields $W$, such
as $Tr[Z^{J_+}W^{J_-}]|0\rangle_{FT}$ will contain all the needed
quantum numbers. There is, however, an ambiguity in the  ordering
structure that should account for the $m_1,\ m_2$ integers in formula
(\ref{etot}).

Next, we discuss the problem of the sigma-model corrections.  Up to
now, we have ignored them, pretending the ground state energy of our
strings to be the classical value $E=m_0J$.  How do we interpret the
corrections in (\ref{etotpoint}) and (\ref{etot})?  A natural
suggestion is a renormalization of the mass of the
constituents. Basically one can think of it as a strong coupling
renormalization, that is, the value $m_0$ is the ``bare'' one in the
strict $\lambda=\infty$ limit, while for finite (but large) $\lambda$
this is corrected as in (\ref{etotpoint}),
(\ref{etot}). Coincidentally, we learn from section \ref{wil} that the
string tension {\it is} renormalized for finite $\lambda$. These are
the same kind of corrections  that are detected in the  hadronic bound
states.  The ratio between the renormalized scalar masses $m(\lambda)$
and the renormalized  string tension is  \bea
\frac{m^2(\lambda)}{T_{QCD}^{(ren)}}=\frac{6\pi}{\lambda}(1+\frac{c}{\lambda})\
, \eea   with $c > 0$. Then the quantum correction increases the
attitude of these fields to decouple as we lower $\lambda$.
Nevertheless, the dimensionless scalar masses  $\hat
m(\lambda)=m(\lambda)/m_0$  decrease with $\lambda$, consistently with
the fact that the expected one-loop mass  square at weak coupling $
m_{scalars}^2\sim \lambda m_0^2$ is much smaller than $m_0^2$.

In summary, we can adapt to our case all the solitonic solutions
already discussed in the case of $AdS_5\times S^5$ (e.g. see the
review article \cite{tsey})  which correspond to strings at rest in
$AdS_5$ and spinning  on a $S^3\subset S^5$. In particular, we have
considered the explicit case of multi-spinning folded and circular
strings which are particular examples of the solutions called
``regular'' in  \cite{tsey}. These solutions admit a regular expansion
in the 't Hooft coupling of the dual ${\cal N}=4$ SYM theory and it is
believed that their quantum corrections to the classical energy are
subleading in the infinite $J$ limit. In confining backgrounds  we
have that one-loop corrections  are not in general subleading but our
results suggest that at one-loop the general energy-spin relation for
these ``regular'' solutions (including the string excitations on the
pp-wave)  admit an expansion in $\lambda/J \ll 1$ of the form \bea
E=m(\lambda)J\Big[1+a_1\frac{\lambda^2}{J^2}+a_2\Big(\frac{\lambda^2}{J^2}\Big)^2+\ldots\Big]\
, \eea where $a_1,a_2,\ldots$ are given by the classical energy-spin
relation while the one-loop quantum effects can be reabsorbed in a
redefinition of the mass of the single hadron constituent
$m(\lambda)$.   Let us observe  that, differently from what happens in
$AdS_5\times S^5$, in our case the ``regular'' solutions  admit an
expansion  in even powers of $\lambda/J$  (up to a nontrivial
dependence of $m(\lambda)$). It would be interesting to understand
better this structure from the dual gauge theory point of view and if
it is preserved by world-sheet higher order  corrections.

Lastly, we recall that in \cite{tomilu} it was speculated that, while
the sigma-model correction could be connected with the renormalization
of the mass $m_0$ of the single constituents (but also accounting for
the mean field of the other constituents), the classical string
dependence in (\ref{etot}) could be related to the correction in the
binding energy in the chain of constituents of the hadron, due to some
mixing between hadrons with different internal structure.  This would
explain why in the hadron built up by only one type of scalar, and so
with only one possible structure, those corrections are not present.

\subsection{Universality of the results}

Let us make some simple observations about the generality of the
results presented in this paper.  The universality of the properties
of the hadrons studied in sections \ref{int}, \ref{hadr}, first
advocated in \cite{gpss}, was extensively discussed in \cite{gfacl},
and we refer the interested reader to those papers.  We concentrate
instead in the results for the Wilson loop and the Regge trajectory.
Apart from numerical coefficients, these calculations are completely
determined by the confining nature (and smoothness) of the background.
Namely, the spectrum of string fluctuations is dictated by the fact
that the metric in the IR is of the type $R^{1,p}\times S^q \times
R^{9-p-q}$, with a non-vanishing $g_{00}$ component.  Crucially, all
the known confining backgrounds are of this form in the IR.  The
actual value of $g_{00}$ (up to quadratic order in the radius) and the
values of $p,\ q$ only affect the final results via the numerical
coefficients in formulae like (\ref{eloopfin}), (\ref{rt}).  As such,
those formulae are expected to be the same, modulo the coefficients,
in all the confining duals.  We then state that the energy of the
rectangular Wilson loop and of the Regge trajectory for glueballs, up
to the first sigma-model correction, are given by the universal
formulae \be\label{relfin1} E_{Wilson}=T_{QCD}\Big(1- {w_1\over
\lambda}\Big)L-\frac{w_2}{L}\ , \ee \be\label{relfin2} J=\alpha'_{adj}
\Big(1-{r_1\over \lambda}\Big)[E_{Regge} - r_0]^2,  \ee with
coefficients $w_1$, $w_2$ (both positive), $r_1$, $r_0$ and bare value
of $T_{QCD}$ (and so $\alpha'_{adj}$) depending on the specific
background.  The first formula is reliable, to this order, up to
exponentially suppressed (in $L$) terms, while the second one includes
only the first order terms in a series expansion.
 
The Wilson loop gives a renormalization of the string tension and a
L\"uscher term.  The non-linear glueball Regge trajectory generically
has a renormalization of the string tension and positive intercept.
The relations (\ref{relfin1},\ref{relfin2}) can be thought of as  the
strong coupling expansions for the physical quantities $E_{Wilson},\
E_{Regge}$.

\section*{Acknowledgments}  
We would like to thank N. Brambilla, P. Fileviez Perez, A. Ozpineci,
J. Sonnenschein and D. Vaman for comments and suggestions.  F. Bigazzi
is partially supported by INFN.  A. L. Cotrone is partially supported
by EC Excellence Grant MEXT-CT-2003-509661.  L. Martucci is partially
supported by the  European Commission RTN program HPRN-CT-2000-00131.
L. Pando Zayas is supported in part by the US Department of Energy
under grant DE-FG02-95ER40899.
 
\appendix

\section{Review of Witten's model} \label{revw} 

Let us briefly review the model proposed in \cite{wqcd} as a possible
IIA string dual of large $N$ Yang-Mills theory in four dimensions.
The background is generated by $N$ D4-branes wrapped on a circle where
antiperiodic boundary conditions for the fermions are assumed. This
breaks supersymmetry explicitly and the low energy gauge theory on the
D4 worldvolume turns out to contain a nonsupersymmetric YM theory in
4-d. As usual, in the supergravity approximation we cannot really
decouple the pure gauge theory from the other modes, so in this limit
only a confining theory sitting, presumably, in the same universality
class as pure YM can be really examined.
  
It is useful to present the 11-d origin of the solution since this
will help us understand it in a wider range of parameters.  The
starting point is the (rescaled, large mass)
Schwarzschild-$AdS_7\times S^4$ solution \bea ds^2_{11}&=&
h(\rho)d\tau^2 + h^{-1}(\rho)d\rho^2 +\rho^2\sum_{i=1}^5dx_idx_i +
\frac{b^2}{4}d\Omega_4^2\ , \nonumber \\ &&
h(\rho)=\frac{\rho^2}{b^2}(1-\frac{b^6}{\rho^6})\ , \qquad\quad
b=2{\alpha'}^{1/2}(\pi N)^{1/3}\ , \eea   where $N>>1$ is the number
of $M5$ branes, wrapped along the circle parameterized by $\tau$,
generating the solution. The background also includes a four-form
field strength which we will write in the following. The coordinate
$\tau$ has periodicity $\delta\tau=2\pi b/3$: this ensures regularity
in the $\rho\to b$ limit. It is possible to  use a normal angular
variable $\theta$ of period $2\pi$ instead of $\tau$, by  changing
coordinates as $\tau= (b/3)\theta$. The fermionic boundary conditions
in going around the $\theta$ circle are taken to be antiperiodic, this
ensures a breaking of supersymmetry on the worldvolume. The
coordinates $\rho$ and $\tau$ have dimension of a length, while the
$x_i$ are dimensionless.
  
Now let us wrap the branes on a second cycle orthogonal to the first
one and parameterized by the angular variable $\beta = 3Nx_5/\lambda$,
where $\lambda$  measures the ratio between the radii of the $\theta$
and $\beta$ cycles at infinity  \be    \frac{R_{\beta}}{R_{\theta}}=
\frac{\lambda}{N}\ .   \ee      From the worldvolume point of view the
wrapping procedure can be understood as follows. One starts with $N$
M5-branes, whose near horizon geometry is $AdS_7\times S^4$ and whose
worldvolume field theory is a six dimensional conformal gauge theory
with $(0,2)$ supersymmetry.  Wrapping the branes along the $\beta$
cycle of radius $R_\beta$ without breaking supersymmetry, one gets a
5-d $SU(N)$ gauge theory with coupling $g_5^2\sim R_\beta$ at low
energy. Then a second compactification is taken, along the $\theta$
circle. The low energy worldvolume theory will then appear four
dimensional and its gauge coupling will go as $g_{YM}^2\sim
R_{\beta}/R_{\theta}$.   If one takes, on the direct product of the
two circles, a supersymmetry-preserving spin structure, the four
dimensional theory will be ${\cal N}=4$ $SU(N)$ SYM. Otherwise, after
imposing antiperiodic boundary conditions to the fermions along the
$\theta$ circle, the low energy theory becomes the pure non
supersymmetric $SU(N)$ YM.    In fact, the fermionic degrees of
freedom get immediately a mass proportional to the inverse radius,
while the would be  massless scalars obtain mass at one-loop level.
The usual limits $N\gg 1$,  $g_{YM}\rightarrow 0$,
$\lambda=g_{YM}^2N=\rm{fixed}$, are taken.
  
After the whole wrapping is performed, the eleven dimensional metric
reads  \bea    ds^2_{11}&=& {\rho^2\over9}(1-{b^6\over
\rho^6})d\theta^2 + {b^2\over\rho^2}(1-{b^6\over \rho^6})^{-1}d\rho^2
+\rho^2\sum_{i=1}^4dx_idx_i + {\rho^2\lambda^2\over9N^2}d\beta^2
+{b^2\over4}d\Omega_4^2 \nonumber \\   &=& e^{-2\Phi/3}ds^2_{10} +
l_s^2e^{4\Phi/3}d\beta^2\ ,  \eea    where $l_s$ is the string length
$l_s^2= {\alpha'}$ and the ten dimensional metric can be read as the
one generated by $N$ D4-branes in type IIA wrapping the $\theta$
circle. From the previous expression, the ten dimensional string frame
metric and dilaton read  \bea    ds_{10}^2&=& {\lambda\rho\over
3Nl_s}\left[{\rho^2\over9}(1-{b^6\over\rho^6})d\theta^2 +
{b^2\over\rho^2}(1-{b^6\over\rho^6})^{-1}d\rho^2
+\rho^2\sum_{i=1}^4dx_idx_i + {b^2\over4}d\Omega_4^2\right]\ ,
\nonumber \\     e^{2\Phi/3}&=& g_s{\lambda\rho\over3Nl_s}\ .   \eea
The parameter $g_s$ is the string coupling in the ``unwrapped'' D4
metric at infinity.  In order to rewrite the whole background in a
more compact form let us define  \be
\label{definits}   
R={b\over2}=(\pi Ng_s)^{1\over3}{\alpha'}^{1\over2}\ , \qquad {u\over
R}= {\rho^2\lambda^2\over 9N^2l_s^2}\ , \qquad u_0={b^3\lambda^2\over
18N^2l_s^2}\ .   \ee     The ten-dimensional string frame metric and
dilaton are thus given by  \bea
\label{defnsdue} 
ds^2&=&({u\over R})^{3/2} (\eta_{\mu\nu}dx^\mu dx^\nu + {4R^3\over
9u_0}f(u)d\theta^2)+ ({R\over u})^{3/2}{du^2\over f(u)}
+R^{3/2}u^{1/2}d\Omega_4^2\ ,\nonumber \\     f(u)&=&1-{u_0^3\over
u^3}\ ,  \nonumber \\     e^\Phi&=&g_s{u^{3/4}\over R^{3/4}}\ ,  \eea
where $u$ has dimensions of length and ranges in $[u_0,\infty)$, and
we shifted to a Minkowski 4-d metric.
  
The background also includes a constant four-form field strength   \be
F_4=3R^3\omega_4\ ,   \ee     where $\omega_4$ is the volume form of
the transverse $S^4$   \be     \int_{S^4}\omega_4={8\pi^2\over3}\ ,
\ee and the normalization of $F_4$ guarantees that the quantization
condition  \be     \int_{S^4}F_4= 8\pi^3{\alpha'}^{3/2}g_s N      \ee
is satisfied.

\subsection{Gauge theory parameters}  
The relation between the $u$ variable and the field theory energy
scale is not known explicitly for the background above. Let us only
suggest that the limit $u\to u_0$ can be interpreted as an IR limit in
the dual field theory. It is in this regime in fact that confinement
was deduced in the string context (through the Wilson loop
analysis). Also notice that in this limit the $g_{00}$ component of
the metric goes to a constant, a necessary condition for confinement
being realized in the dual field theory \cite{sonnensch}.

Let us underline that it is natural to believe that the UV limit in
the dual gauge theory is conversely reached when $u\to\infty$. It is
clear that one of the most difficult problems in the context of the
gauge/gravity correspondence is to find a supergravity description of
the weakly coupled regime of gauge theories. In particular, asymptotic
freedom in supergravity seems to be beyond our current reach. There is
however, an effective vanishing of the coupling in some supergravity
backgrounds. For example, in the case of the Maldacena N\`u\~nez
\cite{mn} model (dual to ${\cal N}=1$ 4-d SYM plus KK matter), the SYM
coupling has been identified  using the DBI action of wrapped
D5-branes \cite{abcpz}. Interestingly, using a possible radius/energy
relation deduced via identification of the dual of the gluino
condensate, it can be shown that the SYM coupling  behaves as a
logarithmic function of the energy scale which is the expected UV
behavior. A logarithmic behavior of the UV beta functions was found
also in the  Klebanov-Strassler background with fractional branes
\cite{KS}.

This unexpected behavior does not occur  in the Witten model.  From
the DBI action for $N$ D4-branes wrapped around the circle
parameterized by $\theta$ in our background (\ref{defnsdue}) one
obtains  \bea   {1\over g^2_{YM}(u)}=\frac{1}{(2\pi)^2l_s} \int
d\theta e^{-\Phi}\sqrt{g_{\theta\theta}}= \frac{1}{3\pi l_s g_s}
\left( \frac{R^3}{u_0}\right)^{\frac12}\sqrt{1-{u_0^3\over u^3}}\ ,
\eea which gives, in the large $u$ limit, the ``geometrical'' constant
value\footnote{This is obtained using $g_{YM}^2=2\pi l_sg_s/R_\theta$
and the relations $R_\theta^2=\frac{4R^3}{9u_0}, R^3=\pi l_s^3g_sN$.}
\bea
\label{gym}   
g_{YM}^2=\frac{3\pi l_s g_s \sqrt{u_0}}{R^\frac{3}{2}} =3\sqrt\pi
\left( \frac{g_s u_0}{l_s N} \right)^\frac12\ .   \eea    So, the
would-be 4-d YM coupling\footnote{We denote as $g_{YM}^2$ and
$\lambda$ the UV couplings, leaving the $u$ dependence in
$g_{YM}^2(u)$ to indicate the running coupling.}  $g_{YM}(u)$
increases while going toward the IR, as expected. However, in the UV
limit it does not go to zero.

The D4-brane map is fairly intricate given that the theory is not
conformal \cite{dp}. It makes sense for us to revise it in the
presence of a thermal circle.  The supergravity regime of validity for
this system spans  IIA and M-theory for appropriate values of the
energies. Let us investigate this topic in more details. As explained
in \cite{dp},  a stack of D4-branes in the Maldacena limit is better
described by starting with a stack of M5-branes wrapped on the
eleventh dimensional circle.  The worldvolume theory is a $(0,2)$ six
dimensional conformal field theory on the M-theory circle.

It is interesting that by looking at the dilaton alone we have a
hierarchy of scales precisely as for the nonthermalized D4 of
\cite{dp}. Recall that the decoupling limit in this case takes the
form  \be  \frac{u}{\alpha'}={\rm fixed}\ , \qquad g_{5}^2=(2\pi)^2
g_s \sqrt{\alpha'} = {\rm fixed}\ , \qquad \alpha' \to 0\ .   \ee
In these quantities the dilaton takes the form  \be e^{\Phi} =
\left(\frac{u^3g_{5}^6}{2^{6}\pi^7\alpha'^3 N}\right)^{1/4}\ .   \ee
For $N^{1/3} \ll g_{5}^2 u/\alpha'$ the appropriate description is in
terms of  11-d supergravity given that in this regime the dilaton is
large.   The decoupling limit for the 11-d theory takes the form
\be  \ell_p\to 0\ , \qquad R_{\beta}=g_s
\sqrt{\alpha'}=g_{5}^2/(2\pi)^2 ={\rm fixed}\ .    \ee
To completely determine the regime of validity of IIA we need two
conditions: small dilaton and small curvatures in string units.  The
Ricci scalar for the 10-d background takes the form
\be {\cal R}=
-\frac{9}{R^{3/2}u^{1/2}}\left(5-\frac{u_0^3}{u^3}\right)\ .   \ee  As
expected for D4-branes we obtain that \cite{dp} $N^{-1}\ll g_{5}^2
u/\alpha' \ll N^{1/3}$. Note that since $u\ge u_0$, the curvature is
always smaller than that of a stack of D4 branes.

We are interested in the regime of validity of the solution in terms
of 4-d parameters.  For this purpose, note that the curvature has its
maximum for $u\rightarrow u_0$, where it is of order   \bea   {\cal
R}\sim \frac{1}{R^{3/2} u_0^{1/2}}\sim \frac{1}{l_s^2 g_{YM}^2N}\ .
\eea   Then the large 't Hooft coupling regime $\lambda\equiv
g_{YM}^2N\gg 1$  is required in order to use the supergravity
approximation.    For what concerns the dilaton, imposing $e^{\Phi}\ll
1$ fixes for $u$ a maximal critical value $u_{crit}=l_s^2 \sqrt{u}_0
N^{\frac13}/R^3 g_{YM}^2$.  Then, $u_0\ll u_{crit}$ means $N^{2/3}\gg
\lambda\gg 1$ and then $g_{YM}^2\ll N^{1/3}$.

Finally, the relevant string parameters can be expressed in terms of
the gauge parameters in the following way \cite{mateos}  \bea
\label{david} R^3=\frac{g_{YM}^2N l_s^2}{3m_0}\
,\qquad g_s=\frac{g_{YM}^2}{3\pi m_0 l_s}\ ,\qquad u_0=\frac13
g_{YM}^2Nm_0l_s^2\ .   \eea

\section{Comment on the solutions at constant radius}\label{appsol}

In this appendix we make some observations on the classical solutions
at constant value of the radius.  Due to the non trivial dependence of
the metric on the radial coordinate $u$, any string with nontrivial
dynamics along the radial direction (i.e. with a non constant
$u(\tau,\sigma)$) has got complicate equations of motions.  In
appendix \ref{appappr} we consider in detail the simplest and most
important example of this kind of non trivial configuration: the
Wilson loop (for other examples of classical string configurations
with a non constant $u(\tau,\sigma)$, see \cite{petkou} and
\cite{pons}).

In this paper we consider  string configurations extending and/or
spinning in the four flat directions or in the internal four-sphere
and satisfying the static gauge $t=k\tau$.  Let us assume that a
general string with this shape lies at a constant $u\neq u_0$.  We
will use the metric in the form given in (\ref{defnsdue}). Since we
are looking for solutions with constant $\theta$ too, from the
Polyakov action in conformal gauge we obtain the following equation of
motion for $u$   \bea \frac{3u^{\frac12}}{2R^{\frac32}}\partial^\alpha
x^\mu \partial_\alpha
x_\mu+\frac{R^{\frac32}}{2u^{\frac12}}G^{S^4}_{IJ}(\chi)\partial^\alpha\chi^I\partial_\alpha\chi^J=0\
, \eea where $\chi^I$ and $G^{S^4}$ are respectively the (arbitrary)
coordinates and the metric for the four sphere.  This equation of
motion must be supplemented with the conformal constraint \bea
\frac{u^{\frac32}}{R^{\frac32}}(\partial_\tau x^\mu \partial_\tau
x_\mu+\partial_\sigma x^\mu \partial_\sigma x_\mu)
+R^{\frac32}u^{\frac12}G^{S^4}_{IJ}(\chi)(\partial_\tau\chi^I\partial_\tau\chi^J+\partial_\sigma\chi^I\partial_\sigma\chi^J)=0\
.  \eea  From this equality we can extract $(\partial_\tau t)^2 $ in
terms of the other coordinates and substituting the result into the
equation of motion we arrive at the condition \bea
\frac{3u^{\frac12}}{R^{\frac32}}\partial_\sigma x^a \partial_\sigma
x^a+\frac{R^{\frac32}}{u^{\frac12}}G^{S^4}_{IJ}(\chi)(\partial_\tau\chi^I\partial_\tau\chi^J+2\partial_\sigma\chi^I\partial_\sigma\chi^J)=0\
.  \eea This implies that, at constant  non-minimal radius $u\neq
u_0$, we can have only solutions with the only non constant
coordinates $x^a=x^a(\tau)$, whose dependence can be solved using the
equation of motion for $x^a$ and correspond to a collapsed string
traveling with the speed of light, i.e.  \bea x^a=x_0^a+n^a\tau\qquad
{\rm with} \qquad n^an^a=k^2\ .  \eea

More general solutions describing strings entirely lying at the
minimal radius $u=u_0$ are instead allowed.  This is easily understood
by looking at the metric in the form (\ref{IRmetric}), since for $r=0$
(and so $u=u_0$) the equation of motion for $r$ is trivially satisfied.

\section{Wilson loop: straight string at non minimal radius}\label{appappr}

Let us consider the more general open string configuration giving the
actual Wilson loop.  This corresponds to the minimal area
configuration spanned by an open string having boundaries at infinite
radius and penetrating for some distance in the bulk, without reaching
the horizon at $u=u_0$, as we will see in a moment.  In order to keep
a clear interpretation in terms of our parameters, we will use the
original coordinates in (\ref{defns}).  Keeping the ``straight'' shape
in $t,\ x^1$, consider the general radial dependence \be
\label{wilgen} t=\tau\ ,\qquad x^1=\sigma\ , \qquad u=u(x^1)\ .  \ee
It is easy to see that this gives a solution of the equations of
motions and that the action reduces to \cite{brand} \be
S=\frac{T}{2\pi}\int dx^1
\sqrt{\frac{u^3}{R^3}+\frac{u'^2}{1-u_0^3/u^3}}\ .  \ee   The equation
for $x^1$ gives the relation \be
\frac{u^3/R^3}{\sqrt{\frac{u^3}{R^3}+\frac{u'^2}{1-u_0^3/u^3}}}= {\rm
const.}\ ,   \ee   from which it follows that \be\label{length}
L=2\int^{\infty}_{u_m}
\frac{u_m^{3/2}R^{3/2}du}{\sqrt{(u^3-u_0^3)(u^3-u_m^3)}}\ , \ee where
we used the fact that the minimum of the radius, $u_m$, is reached, by
symmetry, at $x^1=0$, and that the extremum of $x^1$ is precisely the
string length $L$.   The relation (\ref{length}) gives us $L$ in terms
of an Appell function of $u_0,\  u_m$ evaluated at the extrema.  One
can see that the limit for $u \rightarrow\infty$ is finite, while the
limit for $u \rightarrow u_m$ is an hypergeometric function of $u_0,\
u_m$. We are interested in a semiclassical regime described by large
$L$.

As a first approximation, we can observe the following.  It was shown
in \cite{sonne} that in the large $L$ limit, the classical
configuration of the Wilson loop tends to have a bathtub shape, with
the string coming down from infinity practically straight up to $u_m$,
then becoming suddenly (but smoothly) flat along a special direction
($x^1$ in (\ref{wilgen}))  and then returning back to infinity with
another straight line.  This string spans the loop traveling for a
time $\bar{t}$.  The important point is
that the deviation of this smooth configuration from the rectangular
well is exponentially vanishing with $L$.  Moreover, the lines coming
down from infinity give the infinite quark masses which require
renormalization.  This makes the following straight line approximation
of the above solution \be\label{lineu}   t=\tau\ ,\qquad x^1=\sigma\
,\quad \sigma\in[-\frac{L}{2},\frac{L}{2}]\ , \qquad u=u_m\ , \ee an
accurate one.  Moreover, in the regime $L \rightarrow \infty$,  one
has that $u_m \rightarrow u_0$, as can be verified from
(\ref{length}), where the extremum at $u=\infty$ can be ignored.  More
precisely, (\ref{length}) gives (see \cite{sonnensch} for the analytic
derivation) \be   L\sim
-\frac{2R^{3/2}}{3u_0^{1/2}}\log{\frac{u_m-u_0}{u_0}}\ .  \ee   For
these reasons, the QCD string tension can be simply inferred by
considering the string sitting at $u_0$ and it is classically given by
(\ref{areal}).
  
Analogously, it could be concluded that in this regime it is a good
approximation to go to the $u=u_0$ configuration also for the
quantization procedure.  Let us see why this is the case.
 
Since for large $L$ most of the string is almost straight and reaches
its minimum $u_m$ at its middle point, we can solve the equation of
motion for $u$ in the approximation $y= (u-u_m)/u_0\ll 1$ and
$a_m\equiv u_m/u_0 -1\ll 1$. Up to subleading terms in $y$ or $a_m$,
this reads \bea y^\prime \simeq 3\left(
\frac{u_m}{R^3}\right)^\frac{1}{2}\sqrt{y(y+a_m)}\ , \eea which can be
easily integrated. The result in term of $u$ is the following \bea
u-u_m&\simeq &\frac12(u_m-u_0)\left\{ \cosh\big[  3\left(
\frac{u_m}{R^3}\right)^\frac{1}{2}(\sigma-\sigma_0)\big]-1
\right\}=\cr &\simeq&\frac{u_m}{2}\left\{ \cosh\big[  3\left(
\frac{u_m}{R^3}\right)^\frac{1}{2}(\sigma-\sigma_0)\big]-1 \right\}
e^{-\frac{3L}{2}\left( \frac{u_m}{R^3}\right)^\frac{1}{2}} \ , \eea
where $\sigma_0$ is the middle point of the string,
i.e. $u(\sigma_0)=u_m$. As one can immediately see, $u\rightarrow u_m$
as $L\rightarrow \infty$.  We can now check to what extent  we can use
the above solution. If we take the general constraint
$\sigma-\sigma_0\leq \alpha L/2$ with $\alpha \leq 1$, we obtain the
following upper bound \bea y\sim\leq \frac12 a_m^{1-\alpha}\ .  \eea
Then, in the large $L$ limit, we can use the above approximation for
any $\alpha< 1$. In particular, the approximation is still valid if
\bea 1-\alpha\sim (\log a_m)^{-\beta}\ ,\qquad 0<\beta<1\ , \eea and,
at the leading order in the large $L$ limit, the above approximation
can be extended to the whole string.

We can now turn to the study of the quadratic fluctuations. The only
 nontrivial terms are those for $\theta$ and $u$, while all the other
 modes are massless in the above approximation. Then, we can focus on
 the following terms\footnote{There is also another mass term for $u$,
 but it is exponentially vanishing with $L$.}  of the metric \bea
 ds^2=\ldots +\frac{4R^3}{9u_0}\left(  \frac uR  \right)^\frac32 f(u)
 d\theta^2 + \left(  \frac Ru \right)^\frac32\frac{du^2}{f(u)}+\ldots\
 , \eea which are clearly degenerate in the large $L$ limit. Adding a
 bar to denote the above classical solution as $\bar u(\sigma)$, the
 general dynamical field $u$ can be expanded around $\bar u$. By
 rescaling the fluctuating fields in the following way \be u=\bar u
 +\frac{1}{R^\frac34}\sqrt{\bar u^\frac32 f(\bar u)}\,\zeta\ ,\qquad
 \quad \theta= \frac32\left( \frac{u_0}{R^{\frac32}} \right)^\frac12
 \frac{\chi}{\sqrt{\bar u^\frac32 f(\bar u)}}\ , \ee the  quadratic
 Lagrangian for $\zeta$ and $\chi$ is given by \bea {\cal
 L}_{\zeta,\chi}&\sim&
 \partial_\alpha\chi\partial^\alpha\chi+\partial_\alpha\zeta\partial^\alpha\zeta+\frac{(\bar
 u^{\prime})^2}{2\bar u^\frac32 f(\bar u)}\partial_u^2( u^\frac32
 f(u))_{|u=\bar u}\chi^2 +\\ &&+\left\{ \frac{\bar
 u^{\prime\prime}}{2\bar u^\frac32 f(\bar u)} \partial_u( u^\frac32
 f(u))_{|u=\bar u} -\frac{(\bar u^{\prime})^2}{4(\bar u^\frac32 f(\bar
 u))^2}\big[\partial_u( u^\frac32 f(u))_{|u=\bar u}\big]^2
 \right\}(\chi^2+\zeta^2)\ .\nonumber \eea The first $\chi^2$ term is
 subleading in the large $L$ limit and then the mass for $\chi$ and
 $\zeta$ is the following \bea M^2_{\chi,\zeta}&\simeq& \frac{\bar
 u^{\prime\prime}}{2\bar u^\frac32 f(\bar u)} \partial_u( u^\frac32
 f(u))_{|u=\bar u} -\frac{(\bar u^{\prime})^2}{4(\bar u^\frac32 f(\bar
 u))^2}\big[\partial_u( u^\frac32 f(u))_{|u=\bar u}\big]^2\simeq\cr
 &\simeq& \frac{3\bar u^{\prime\prime}}{2u_m f(\bar u)}-\frac{9(\bar
 u^{\prime})^2}{4u_m^2(f(\bar u))^2}\simeq\cr &\simeq&
 \frac{9u_m}{4R^3}\left\{\frac{2\cosh\big[  3\left(
 \frac{u_m}{R^3}\right)^\frac{1}{2}(\sigma-\sigma_0)\big]}{\cosh\big[
 3\left(
 \frac{u_m}{R^3}\right)^\frac{1}{2}(\sigma-\sigma_0)\big]-1}-\frac{\sinh^2\big[
 3\left(
 \frac{u_m}{R^3}\right)^\frac{1}{2}(\sigma-\sigma_0)\big]}{\big\{\cosh\big[
 3\left(
 \frac{u_m}{R^3}\right)^\frac{1}{2}(\sigma-\sigma_0)\big]-1\big\}^2}
 \right\}=\cr &=&\frac{9u_m}{4R^3} \ .  \eea To summarize, at the
 leading order the string can be considered as an almost straight
 string lying at $u_m$ for almost all its length and with two massive
 modes with constant, $\sigma$-independent mass $M_{\chi,\zeta}$. The
 corresponding frequencies reduce to the ones found in section
 \ref{wbqf} (see eq. (\ref{wilbos1})) in the large $L$
 ($u_m\rightarrow u_0$) limit.

\section{Regge trajectory: anomaly cancellation}\label{appweyl}

In this appendix we show why the quantized configuration of section
\ref{reg} is free of conformal anomalies.  Let us start by observing
how the $\sigma$ dependent mass term in the bosonic action
(\ref{actionb}) can be  reabsorbed as a curvature effect for 2-d
scalar fields with constant mass on a  curved world-sheet with metric
given by the induced metric  $h_{\alpha\beta}=k^2
\cos^2{\sigma}\,\eta_{\alpha\beta}$.  Indeed, (\ref{actionb}) can be
rewritten as \bea
\label{actionb2}
S=-\frac{1}{3\pi}\int d\tau d\sigma\sqrt{-h}
\sum_{a=1,2}\big[h^{\alpha\beta}(\nabla_\alpha y_a \nabla_\beta y_a)
+\frac94(y_a)^2\big]\ ,  \eea   where we have introduced the 2-d
world-sheet covariant derivative $\nabla_\alpha$  to put the action in
an explicitly covariant form.   Then the equations of motion become
\bea\label{boperator} \big[h^{\alpha\beta}\nabla_\alpha \nabla_\beta
-\frac{9}{4}]y_a=0\ .  \eea The two descriptions with flat and curved
world-sheet in (\ref{actionb}) and (\ref{actionb2}), which are related
by a classical  Weyl transformation, can be really considered
equivalent only if the conformal anomaly vanishes, a necessary
condition for the consistency of the theory  which  is also related to
the finiteness of the theory  (and as a consequence, of the one-loop
correction to the energy).

In the same way, let us write the fermionic action (\ref{actionf1}) as
an action for fermions on a curved world-sheet with metric equal to
the  induced metric $h_{\alpha\beta}$.  If we organize the spinors
$\Theta^I$ in eight 2-d Majorana spinors $\psi^i=\psi_i$, the action
(\ref{actionf1}) can be written as \bea S_F&=&\frac{i}{4\pi}\int d\tau
d\sigma\sqrt{-h}\big[h^{\alpha\beta}\bar\psi^i\tau_\alpha
D^{(2)}_\beta\psi_i-\frac34\bar\psi^i  \tilde\gamma_{ij}\psi^j \big]\
, \eea where $\tau_\alpha$ are the curved 2-d gamma matrices and
$D_\alpha^{(2)}$ is the 2-d spinorial covariant derivative.  Then the
equations of motion become \bea h^{\alpha\beta}\tau_\alpha
D^{(2)}_\beta\psi^i-\frac34\tilde\gamma^i{}_j\psi^j=0\ .  \eea These
equations can be squared into the equations \bea\label{foperator}
\big[h^{\alpha\beta}\hat\nabla_\alpha\hat\nabla_\beta-\frac14{\cal
R}^{(2)}-\frac{9}{16}\big]\psi^i\ , \eea where $\hat\nabla_\alpha$ is
the complete covariant derivative, containing both the spin connection
and the Christoffel symbols, and ${\cal R}^{(2)}$ is the 2-d scalar
curvature and is equal to \bea {\cal R}^{(2)}=\frac{2}{k^2\cos^2
\sigma}(1+\tan^2\sigma)\ .  \eea  As we will see in a moment, this
action has the right form in order to combine well with the bosonic
and ghost contributions, giving a finite (at least at one-loop) theory
with vanishing  conformal anomaly.

In fact, following the analysis of \cite{grosstsey} and starting from
the actions with the curved induced metric above, it is possible to
extract the one-loop contribution to the partition function computing
the determinants of the second order operators which enter the bosonic
and  fermionic equations of motions (\ref{boperator}) and
(\ref{foperator}), supplemented by the ghost contribution.  Working in
the Euclidean formulation and regularizing the determinants with an UV
cutoff, it is possible to show in general how these determinants have
quadratically, linearly and logarithmically divergent terms. The
quadratic and linear divergences  trivially drop out in our case due
to the matching between the bosonic and fermionic degrees of freedom.
The logarithmic term is more delicate and can be  expressed in term of
the integral of the following Seeley coefficients for bosons, fermions
and ghosts \bea &&b_{2B}=10\times \frac{{\cal R}^{(2)}}6\,-\sum_B
m_B^2 \ ,\qquad b_{2F}=8\times \frac{{\cal R}^{(2)}}3\,+\sum_F m_F^2\
,\cr &&\qquad\qquad\qquad\quad b_{2gh}=-2\times \frac{{\cal
R}^{(2)}}6\, - {\cal R}^{(2)}\ .  \eea           Since in our case we
have only two massive bosons with equal masses $m_B^2=\frac94 m_0^2$
and eight  massive fermions with equal masses
$m_F^2=\frac{9}{16}m_0^2$  (reintroducing the $m_0$ dependence), we
have that  the mass matching condition  \bea\label{mmc} \sum_B
m_B^2=\sum_F m_F^2\ , \eea implies that the resulting total divergence
coefficient is then the integral of \bea b^{(tot)}_2=3{\cal R}^{(2)}\
, \eea  which is proportional to the Euler character. As explained in
\cite{grosstsey}, this term is exactly canceled  by the cutoff
dependent factors in the conformal Killing vector and/or Teichm\"uller
measure.  Then the theory is finite and since the above Seeley
coefficients are exactly the same ones appearing in the Weyl anomaly,
the theory is also anomaly free at one-loop. This is indeed what we
have verified explicitly in the other computations of the one-loop
energy considered in this paper where the world-sheet induced metric
is constant and the finiteness of the one-loop correction is due to
the same mass matching condition (\ref{mmc}). The same happens  for
the Regge trajectories.  Going to the flat world-sheet metric gauge,
we obtain an analogous ($\sigma$-dependent) mass matching condition
\bea\label{mmc2} \sum_B m_B^2(\sigma)=\sum_F m_F^2(\sigma)\ , \eea
which again checks the UV finiteness of the 2-d theory, as for example
happens for the folded strings considered  in \cite{frolov1}.

\section{Penrose limit}\label{apppenr}

We perform here the Penrose limit of the Witten background along the
great circle of the four-sphere.  Since we are interested in the IR
regime, we will use the metric expanded around $u= u_0$
(\ref{IRmetric}) \be\label{IRmetric2}  ds^2\approx ({u_0\over
R})^{3/2}[1+{3r^2\over 2}](\eta_{\mu\nu}dx^\mu dx^\nu) +
{4\over3}R^{3/2}\sqrt{u_0}(dr^2+ r^2 d\theta^2) + R^{3/2}u_0^{1/2}[1+
{r^2\over 2}]d\Omega_4^2\ ,   \ee We will also use the fact that in
$u=u_0$ the dilaton goes to a constant $e^\Phi\to g_s
u_0^{3/4}/R^{3/4}$.
  
Using the parameterization $\chi,\psi,\phi_-,\phi_+$ introduced in
(\ref{curved}) for the $S^4$,  the null geodesic we want to zoom in is
determined by the following   conditions on the coordinates \bea
t=\phi_+\ ,\qquad x^i=r=\psi=\chi=0\ .  \eea As usual we first shift
the $3+1$ coordinates as\footnote{Clearly this $L$ has nothing to do
with the string length in section \ref{wil}.}  $x^{\mu}\to
(R^{3/4}L/u_0^{3/4})x^{\mu}$ and send $L\to\infty$ while keeping \be
m_0^2 \equiv {L^2\over{{\sqrt u_0}R^{3/2}}}\quad \rm{fixed}.
\label{mzero}  
\ee   This would ensure that, in the limit, we are keeping fixed the
matter field and glueball masses, while taking the string tensions,
proportional to $\alpha'L$, very large.   Notice that $R^{3/2}\approx
\sqrt N$. Also, recall from (\ref{definits}) that
$u_0=b^3\lambda^2/(18N^2l^2)$, so that taking $\lambda$ very large,
namely of order $N$, we get $u_0\approx N$ and $L^2\approx N$ in the
limit. The dilaton in the IR goes like $e^{\Phi_0}\approx g_sN^{1/2}$.
To be consistent with the notations in the rest of the paper, we can
choose $L^2=u_0^{3/2}/R^{3/2}\approx g_{YM}^2\,N$ so that indeed
$m_0^2=u_0/R^3$.
  
In order to perform the Penrose limit we re-shift $x^i\to x^i/L$ and
take   \be   \psi={m_0\over L}v_3\ ,\qquad \chi= {  m_0\over L}y\ ,
\qquad   r={{\sqrt3}m_0\over 2L}\rho\ .   \ee Expanding the whole
metric near the above defined null geodesic gives \bea
ds^2&=&[1+{9m_0^2\rho^2\over8L^2}](-L^2dt^2 + {L^2\over
m_0^2}d\phi_+^2) +dx^idx^i + d\rho^2 + \rho^2d\theta^2 +\nonumber \\
&&+ dy^2 +y^2 d\phi_-^2 -[y^2 + v_3^2 +{3\over4}\rho^2]d\phi_+^2
+dv_3dv_3\ .   \eea   Now, let us introduce the complex variables \be
u=\rho e^{i\theta}\ , \qquad v=ye^{i\phi_-}\ ,   \ee   and the
light-cone coordinates   \be   t=x^+\ , \qquad x^-=
{L^2\over2}(t-{\phi_+\over m_0})\ .   \ee   This way we get the
plane-wave metric \be   ds^2=-4dx^+dx^- - m_0^2[v_3^2+v{\bar v} +
{3\over4}u{\bar u}]dx^+dx^+ + dx^idx^i + dud{\bar u} + dvd{\bar v} +
dv_3^2\ .  \ee

Written explicitly in terms of the angular variables introduced above
the four-form RR field strength is   \be   F_4=
{3R^3}\cos^3\psi\sin\chi\cos\chi d\psi\wedge d\chi\wedge d\phi_-\wedge
d\phi_+\ .  \ee   After taking the Penrose limit we get   \be   F_4 =
{3i\over2u_0^{3/4}}R^{3/4}m_0\ dx^+\wedge dv_3\wedge dv\wedge d{\bar
v}\ .   \ee   Notice that this object still scales in the limit: in
fact $R^{3/4}/u_0^{3/4}\approx N^{-1/2}$. In the very same way as in
the D2 and fractional D2 case of \cite{gfacl}, the correct non scaling
object is   \be   e^{\Phi_0}F_4 = {3i\over2} m_0 dx^+\wedge dv_3\wedge
dv\wedge d{\bar v}\ .   \ee   It is now easy to check that the only
non-trivial equation of motion for the whole pp-wave background at
hand   \be   R_{++}= {1\over12}e^{2\Phi_0}F_{+abc}{\bar F}_{+abc} \ee
is satisfied.

\end{document}